\documentclass[aps,prb,twocolumn,showpacs,superscriptaddress]{revtex4}
\usepackage{graphicx}
\usepackage{bm}

\usepackage{graphicx}
\usepackage{bm}

\newcommand{\beq}{\begin{equation}}
\newcommand{\eeq}{\end{equation}}

\begin{document}

\title{Topology of the Fermi Surface Beyond the Quantum
Critical Point}
\author{V.~A.~Khodel}
\affiliation{Russian Research Centre Kurchatov
Institute, Moscow, 123182, Russia}
\affiliation{McDonnell Center for the Space Sciences \&
Department of Physics, Washington University,
St.~Louis, MO 63130, USA}
\author{J.~W.~Clark}

\affiliation{McDonnell Center for the Space Sciences \&
Department of Physics, Washington University,
St.~Louis, MO 63130, USA}
\affiliation{Complexo Interdisciplinar, Centro de
Mathem\'atica e Aplica\c{c}\~{o}es Fundamentals\\
University of Lisbon, 1649-003 Lisbon, Portugal}
\affiliation{Departamento de F\'isica, Instituto Superior T\'ecnico,
1096 Lisbon, Portugal}
\author{M.~V.~Zverev}
\affiliation{Russian Research Centre Kurchatov
Institute, Moscow, 123182, Russia}
\date{\today}
\begin{abstract}
We examine the nature of phase transitions occurring in
strongly correlated Fermi systems at the quantum critical
point (QCP) associated with a divergent effective mass.
Conventional scenarios for the QCP involving collective degrees
of freedom are shown to have serious shortcomings.  Working
within the original Landau quasiparticle picture, we propose
an alternative topological scenario for the QCP, in systems
that obey standard Fermi liquid (FL) theory in advance of
the QCP.  Applying the technique of Poincar\'e mapping, we
analyze the sequence of iterative maps generated by the Landau
equation for the single-particle
spectrum at zero temperature.  It is demonstrated that the
Fermi surface is subject to rearrangement beyond the QCP.
If the sequence of maps converges, a multi-connected Fermi
surface is formed.  If it fails to converge, the Fermi surface
swells into a volume that provides a measure of entropy associated
with formation of an exceptional state of the system characterized
by partial occupation of single-particle states and dispersion
of their spectrum proportional to temperature.  Based on this dual
scenario, the thermodynamics of Fermi systems beyond the QCP
exhibits striking departures from the predictions of standard
FL theory.  Mechanisms for the release of the entropy excess
of the exceptional state are discussed.

\end{abstract}

\pacs{
71.10.Hf, 
71.27.+a,  
71.10.Ay  
}
\maketitle

\section{Introduction}

The Landau quasiparticle pattern \cite{lan} of low-temperature
phenomena in Fermi systems is universally recognized as a cornerstone
of condensed-matter theory. Although Landau's Fermi liquid (FL)
theory was originally formulated for liquid $^3$He,
it is a quirk of fate that discrepancies between theoretical
predictions of the theory and experimental data on
the low-temperature behavior of key thermodynamic properties
first came to light in this system.

In the three-dimensional (3D) liquid, deviations of the spin
susceptibility $\chi(T)$ and specific-heat ratio $C(T)/T$ from
the predicted constant behavior, rather small at very low
temperatures $T$, increase with
$T$. In 2D liquid $^3$He, however, departures from the predictions
of FL theory become more pronounced as the temperature is
{\it lowered} once the density $\rho$ reaches a critical
region where the FL effective mass $M^*(\rho)$ is greatly
enhanced,\cite{godfrin,saunders,sscience} the values of $M^*$
having been extracted from the specific heat data via
the FL formula $C_{\mbox{\scriptsize FL}}(T)=Tp_FM^*/3$.
This contrary behavior rules out damping of single-particle
excitations as the cause of the observed failure of
FL theory in 2D liquid $^3$He, since damping effects
decrease and vanish at $T\to 0$.  The experimental
data on the spin susceptibility $\chi(T)$ of $^3$He films
present a thorny challenge of interpretation. In agreement with the
Curie law but against the FL predictions, the product $\chi(T) T$
fails to show vanishing behavior as very low values of $T$ are reached
in the measurements. In the critical density region, the value of
this product {\it gradually increases} with increasing $\rho$, as
if a fraction of spins of $^3$He atoms becomes localized and
coexists with the liquid part of the system. Analogous behavior
has been observed for 2D electron gas.\cite{reznikov}
Non-Fermi-liquid (NFL) behavior also exhibits itself in properties
of strongly correlated electron systems of heavy-fermion
metals.\cite{stewart,takahashi,steglich,gegenwart,custers,steglich3,budko,yuan1,
yuan2,paglione}

Various theories, generally invoking critical
fluctuations associated with second-order phase transitions,
\cite{hertz,millis} have been proposed to explain NFL
behavior in strongly correlated Fermi systems beyond the
quantum critical point (QCP).  At a QCP, which by definition
occurs at zero temperature, the Landau state becomes unstable due
to a divergence of the effective mass $M^*$
(cf.~Ref.~\onlinecite{coleman1} and works cited therein).

On the weakly correlated side of the QCP, the properties of
systems of interest are described within the {\it standard}
FL theory, in which the Fermi liquid is treated as a {\it gas}
of interacting quasiparticles with a spectrum of single-particle
excitations given by $\epsilon(p)=p_F(p-p_F)/M^*$.  Since
this spectrum loses its meaning when $M^*$ diverges,
the standard FL theory fails at the QCP and beyond.
Bearing in mind the fundamental role played by the Landau
approach in modern condensed matter physics, there is
ample incentive to investigate the situation beyond the
QCP within the {\it original} Landau quasiparticle
formalism,\cite{lan} free of shortcomings of the
standard quasiparticle picture.  We will demonstrate that if
the strength of the effective interaction between quasiparticles
reaches a critical level, a {\it topological} phase transition
occurs, and the properties of the system change dramatically.

This paper is organized as follows.
In Sec.~II we call attention to certain basic flaws of the
conventional {\it collective} scenario for the QCP that envisions
a ``fatal'' breakdown of the quasiparticle picture when the
quasiparticle weight $z$ vanishes at a second-order phase
transition. We then describe an alternative {\it topological}
scenario for the QCP, in which the quasiparticle group velocity at
the Fermi surface changes sign while {\it the $z$-factor stays
finite}.  In Sec.~III we apply the concept and technique of
Poincar\'e mapping, widely used in the
analysis of nonlinear phenomena, to explore the
structure of approximants (iterates) generated by iteration of the
nonlinear integral equation \cite{lan} for the single-particle
spectrum $\epsilon(p)$
and momentum distribution $n(p)$
at zero temperature.
Beyond the QCP, the iteration procedure either converges, in
which case the Fermi surfaces becomes multi-connected;
or it does not.  In the latter case, which applies for
systems with long-range effective forces, the patterns
of successive iterates for $\epsilon(p)$ and $n(p)$
acquire chaotic features.  A special procedure is introduced
for averaging over sequences of iterates, in such a way
as to suppress these chaotic features.   In Sec.~IV, we
demonstrate that the resulting averaged single-particle
energies and occupation numbers coincide with those
belonging to a state with a fermion condensate--an exceptional
type of ground state possessing a distinctive topological
structure.
Sec.~V presents results from numerical calculations based on the Landau
equation for the spectrum $\epsilon(p)$ at finite temperatures
beyond the points of the topological phase transitions.
The relevance of the topological scenario to the real
experimental situation in Fermi liquids beyond the QCP is
discussed in Sec.~VI.  Extraordinary aspects of the NFL thermodynamics
of these systems---Curie-Weiss behavior of the spin susceptibility
and the role of the entropy excess
associated with the fermion condensate---are addressed in Sec.~VII.
In Sec.~VII, we analyze different possibilities for release of
this excess entropy, which may be responsible for the diversity
of quantum phase transitions appearing in the phase diagrams of
strongly correlated Fermi systems beyond the QCP.   The paper is
concluded in Sec.~IX with a summary of key developments and
findings, and with remarks on the true scope of Landau theory.

\section{Two different scenarios for the quantum critical point}
A dominant activity in condensed-matter physics during the last
decade has been the investigation of quantum phase transitions,
occurring at extremely low temperatures in strongly correlated
Fermi systems. Imposition of pressure or magnetic
fields allows one to push the transition temperature toward zero,
producing a QCP associated with divergence of the density of states, or
equivalently, the effective mass $M^*(\rho)$ at a critical density
$\rho_{\infty}$.  In homogeneous nonsuperfluid Fermi systems, the ratio
of $M^*$ to the bare mass $M$ is determined by the textbook formula
\begin{equation}
{M\over
M^*}=z\left[1+\left({\partial\Sigma(p,\varepsilon)\over\partial\epsilon^0_p}
\right)_0\right] \  ,
\label{meffz}
\end{equation}
where $\epsilon^0_p=p^2/2M-\mu$ is the bare quasiparticle energy
measured relative to the chemical potential,
$\Sigma$ represents the mass operator, and the quasiparticle
weight $z$ in the single-particle state is given by
$z=[1-\left(\partial\Sigma(p,\varepsilon)/
\partial\varepsilon\right)_0]^{-1}$.
Here and henceforth, the subscript $0$ indicates that the
quantity in question is evaluated at the Fermi surface.

Since the Feynman-Dyson era, it has been a truism that the effects
of the two factors in Eq.~(\ref{meffz}), associated respectively
with (i) the energy dependence of the mass operator (or self-energy)
$\Sigma(p,\varepsilon)$ and (ii) its momentum dependence, cannot
be separated from each other {\it based only on measurements of
thermodynamic and transport properties}.  Importantly, at the
quantum critical point these factors express themselves differently
in different scenarios. In a conventional {\it collective}
scenario for the QCP, the energy dependence of $\Sigma$ plays a
decisive role.  ``Quasiparticles get heavy and die," \cite{coleman1}
since critical fluctuations destroy the quasiparticle picture,
causing the quasiparticle weight $z$ to vanish at the transition
point.\cite{doniach,dyugaev,steglichrev}

By contrast, in a {\it topological} scenario\cite{khodel} for the QCP,
which is associated with a change of sign of the quasiparticle group
velocity $v_F=(d\epsilon(p)/dp)_0$ appearing in Eq.~(\ref{meffz}),
the momentum dependence of the mass operator evidently assumes
a key role, when we note that $v_F$ is proportional to the
sum $1+\left(\partial\Sigma(p,\varepsilon=0)/
\partial\epsilon^0_p\right)_0$.
In this scenario, nothing catastrophic happens beyond the critical
point where $v_F$ reverses sign: the original quasiparticle
picture, advanced by Landau in the first article devoted to theory
of Fermi liquids,\cite{lan} holds on both sides of the QCP,
since the $z$-factor remains finite.

\subsection{An original Landau quasiparticle pattern}

We recall that the heart of the Landau quasiparticle
picture is the postulate that there exists a one-to-one
correspondence between the totality of real, decaying
single-particle excitations of the actual Fermi liquid and a
system of immortal interacting quasiparticles. Two features
specify the latter system. First, the number of quasiparticles is
equal to the given number of particles (the so-called
Landau-Luttinger theorem).  This condition is expressed as
\beq
2\int n(p)\,d\upsilon=\rho  \  ,
\label{part}
\eeq
where $n(p)$ is the quasiparticle momentum distribution, the
factor 2 comes from summation over the spin projections,
and $d\upsilon$ is a volume element in momentum space.

Second, the entropy $S$, given by the combinatorial expression
\begin{equation}
S(T)=-2\int \left[n(p)\ln n(p) +
        (1{-}n(p))\ln(1{-}n(p)) \right]
d\upsilon \ ,
\label{entr}
\end{equation}
based on the quasiparticle picture, coincides with the entropy
of the actual system. Treating the ground-state energy $E_0$
as a functional of $n(p)$, Landau derived the formula
\beq
n(p,T)=\left[1+e^{\epsilon(p)/T}\right]^{-1} \  \  , \label{dist}
\eeq
which has an obvious (but misleading) resemblance to the
Fermi-Dirac formula for the momentum distribution of an ideal
Fermi gas. In contrast to the ideal-gas case, the quasiparticle energy
$\epsilon(p)=\delta E_0/\delta n(p)-\mu$ itself must be treated as
a functional of $n(p)$.

Another fundamental relation, which follows from the Galilean
invariance of the Hamiltonian of the problem, connects the
group velocity $\partial\epsilon(p)/\partial{\bf p}$
of the quasiparticles to their momentum distribution
through \cite{lan,lanl,trio}
\beq
{\partial\epsilon(p)\over\partial {\bf p}} =
{{\bf p}\over M} + 2\int\! f({\bf p},{\bf p_1})\,
{\partial n(p_1)\over\partial {\bf p_1}}\, d\upsilon_1  \  .
\label{lansp}
\eeq
The interaction function $f({\bf p},{\bf p}_1)$ appearing
in this relation is the product of $z^2$ and the scalar part
of the scattering amplitude
$\Gamma^{\omega}$. In turn, $\Gamma^{\omega}$ is the
$\omega$-limit of the scattering amplitude $\Gamma$ of two
particles whose energies and incoming momenta ${\bf p}_1,{\bf
p}_2$ lie on the Fermi surface, with scattering angle
$\cos\theta={\bf p}_1{\bf p}_2/p^2_F$ and the 4-momentum transfer
$({\bf q},\omega)$ approaching zero such that $q/\omega\to 0$.

Landau (see formula (4) in Ref.~\onlinecite{lan}) supposed that
solutions of Eq.~(\ref{lansp}) always arrange themselves in
such a way that at $T=0$, the quasiparticle group velocity $v_F$
maintains a positive value, implying that the quasiparticle
momentum distribution takes the Fermi-step form
$n(p,T=0)=n_F(p)=\theta(p-p_F)$. If this supposition holds,
implementation of the
original quasiparticle picture is greatly facilitated: properties
of any Fermi liquid coincide with those of {\it a gas} of
interacting quasiparticles.

The failure of this assumption in strongly correlated Fermi
systems, established first in microscopic calculations of
Refs.~\onlinecite{ks1,zjetp}, can be seen from the analysis of
Eq.~(\ref{lansp}) itself. Consider for example homogeneous
fermionic matter in 3D. Upon setting $T=0$ and $p=p_F$ in
Eq.~(\ref{lansp}) and denoting the first harmonic of the
interaction function by $f_1$ we find
\beq
v_F={p_F\over M}\left(1- {1\over 3}F^0_1\right)\  \
, \label{group} \eeq
having introduced  the dimensionless parameter $F^0_1=p_FMf_1/\pi^2$.
Eq.~(\ref{group}) is easily rewritten in the FL form \cite{lan,lanl,trio}
\beq
{M\over M^*}=1-{1\over 3} {p_FM\over \pi^2}f_1\equiv 1-{1\over 3} F^0_1
  \ .
  \label{meffl}
\eeq
Hence the inequality $v_F>0$ is violated at the critical
density $\rho_{\infty}$ where $F^0_1(\rho_{\infty})=3$.

It is instructive to rewrite Eq.~(\ref{meffl}) in terms of the
$k$-limit of the dimensionless scattering amplitude
$\nu\Gamma^k=A+B {\bm \sigma}_1{\bm \sigma}_2$, where $\nu
=z^2p_FM^*/\pi^2$ is the quasiparticle density of states.
Noting the connection\cite{lan,lanl,trio}
$A_1=F_1/( 1+F_1/3)$ of $A_1$ to  $F_1=p_FM^*f_1/\pi^2$,
simple algebra based on Eq.~(\ref{meffl}) then yields
\beq
{M\over M^*}=1-{1\over 3} A_1  \  .
\label{meffm}
\eeq
Clearly then, one must have
\beq
A_1(\rho_{\infty})=F^0_1(\rho_{\infty}) = 3
\label{critf}
\eeq
at the density $\rho_{\infty}$ where the effective mass
diverges.

We would like to emphasize that the critical density $\rho_{\infty}$
has no relation to the density associated with violation of the
Pomeranchuk condition for stability against dipolar deformation of
the Fermi surface. In  the latter situation, the critical
value $M^*_{1c}$ of the effective mass $M^*$ is found from
the condition \cite{lan,lanl,trio}
\beq
1+f_{1c}{p_FM^*_{1c}\over 3\pi^2}=0 \ ,
\eeq
which, according to Eq.~(\ref{meffl}),  may be recast
as $(1-F^0_{1c}/3)^{-1}=M^*_{1c}/M=0$. Quite evidently, this
situation is not relevant to the QCP.

\subsection{Critique of the conventional QCP scenario}
In this subsection, we show that the collective scenario for the
QCP encounters difficulties when the wave vector $k_c$ specifying
the spectrum of the critical fluctuations has a nonzero value.
This result casts doubt on the standard collective QCP scenario,
since
the assumption of a finite critical wave vector number $k_c$ is aligned
prevalent ideas.

\subsubsection{Homogeneous matter}
We first address the homogeneous case, and, for the sake
of specificity, we focus on density fluctuations\cite{cgy} as the
seed for the transition.  The requirement of antisymmetry of the
amplitude $\Gamma$ with respect to interchange of the momenta and
spins of the colliding particles leads to the
relation\cite{dyugaev}
\beq
  A({\bf p}_1,{\bf p}_2,{\bf k},\omega=0;\rho\to \rho_c)=
  -D({\bf k})+{1\over 2} D({\bf p}_1-{\bf p}_2+{\bf k})
  \label{dyug} \,,
\eeq
where
\beq
D(k\to k_c,\omega=0;\rho\to \rho_c)={g\over \xi^{-2}(\rho)
+(k-k_c)^2}  \  , \label{crit}
\eeq
with $g>0$ and the correlation length $\xi(\rho)$ divergent at $\rho=\rho_c$.

To explicate difficulties encountered by the standard scenario for
the QCP in homogeneous matter, let us calculate harmonics
$A_k(\rho)$ of the amplitude $A(p_F,p_F,\cos\theta)$ from
Eqs.~(\ref{dyug}) and (\ref{crit}). In particular, we obtain
\beq
A_0(\rho{\to}\rho_c)=g{\pi\over 2}{k_c\xi(\rho)\over p^2_F}, \;
A_1(\rho{\to}\rho_c)=g{3\pi\over 2} {k_c\xi(\rho)\over
p^2_F}\cos\theta_0    \   .
\label{harm}
\eeq
The sign of $A_1(\rho\to \rho_c)$, which coincides with the sign of
$\cos\theta_0=1-k^2_c/2p^2_F$, turns out to be {\it negative} at
$k_c>p_F\sqrt{2}$. According to Eq.~(\ref{meffm}), this means that
at the second-order phase transition, the ratio $M^*(\rho_c)/M$
must be less than unity.  We are then forced to conclude that
the densities $\rho_c$ and $\rho_{\infty}$ {\it cannot coincide}.
Although the $z$-factor does vanish at the density $\rho_c$
due to the divergence of the derivative
$\left(\partial\Sigma(p,\varepsilon)/\partial\varepsilon\right)_0$,
the effective mass $M^*$ remains finite, since the derivative
$\left(\partial\Sigma(p,\varepsilon)/\partial\epsilon^0_p\right)_0$
diverges at the QCP as well.\cite{khodel}

Moreover, it is readily demonstrated that at positive values of
$\cos\theta_0$ below unity, vanishing of the $z$-factor is {\it
incompatible} with divergence of $M^*$. Indeed, as seen from
Eq.~(\ref{harm}), the harmonics $A_0(\rho_c)$ and $A_1(\rho_c)$
are related to each other by  $A_0(\rho_c)=A_1(
\rho_c)/(3\cos\theta_0)$. If  $M^*(\rho_c)$ were infinite, then
according to Eq.~(\ref{meffm}), $A_1(\rho_c)$  would equal  3,
while $A_0(\rho_c)=1/\cos\theta_0>1$. On the other hand, the basic
FL connection $A_0=F_0/(1+F_0)$ implies that $A_0\leq 1$, provided
the Landau state is stable. Thus in the conventional scenario, the
QCP cannot be reached without violating the stability
conditions for the Landau state. The same is true for critical
spin fluctuations with nonvanishing critical wave number. One must
conclude that the system, originally obeying FL theory, undergoes
a first-order phase transition upon approaching the QCP, as in
the case of 3D liquid $^3$He.

Based on these considerations we infer that for homogeneous
matter, the list of possible second-order phase transitions
compatible with the divergence of the effective mass $M^*$
includes only long-wave transitions.  These are associated
with some $L-$deformation of the Fermi surface such that one of
the two
Pomeranchuk stability conditions,\cite{trio} e.g.,
$1+F_L/(2L+1)>0$, is violated.

\subsubsection{Anisotropic systems}
We turn now to systems, typified by heavy-fermion metals,
whose Fermi surface is anisotropic.  In the energy region
$|\epsilon|\simeq T$ most relevant to the thermodynamic
and transport properties of all Fermi systems, including
anisotropic examples, the quasiparticle group velocity
${\bf v}=\partial\epsilon/\partial{\bf p}$ consists
only of the component $v_n$ normal to the Fermi surface where
$\epsilon=0$; hence it depends only on the corresponding
momentum component $p_n$.
In homogeneous matter, the value of the velocity is the same
at any point of the Fermi surface, whereas in the anisotropic case, it
depends on the observation point.  Even so, the strategy used
below to investigate topological phase transitions in anisotropic
systems is much like that applied above in the analysis of such
transitions in homogeneous matter.

To deal with electronic systems of solids, Eq.~(\ref{meffz}) is
replaced by
\beq
v_n({\bf p})= \left({\partial\epsilon({\bf p})\over \partial
p_n}\right)_0= \left({\partial \epsilon^0( {\bf p})\over \partial
p_n}\right)_0 z \left[1+\left({\partial\Sigma( {\bf
p},\varepsilon)\over\partial  \epsilon^0({\bf p})
}\right)_0\right]\,,
\eeq
where $\epsilon^0({\bf p})$ is the ``bare'' electron spectrum,
evaluated so as to account for the external field due to the
lattice, while the electron mass operator $\Sigma$ includes
all the electron-electron interaction effects.

Violation of Galilean invariance in solids results in a change
of Eq.~(\ref{lansp}) stemming from the FL relation between
the $k$- and $\omega$-limits of the vertex ${\cal T}({\bf p})$,
because the Pitaevskii identity\cite{trio}
$z{\cal T}^{\omega}( {\bf p})={\bf p}$ used in deriving this
equation is no longer valid.  Accordingly, we arrive at
the generalized version
\beq
 \left({\partial \epsilon( {\bf p})\over \partial  p_n}\right)_0=
z\left({\cal T}^{\omega}(  p_n)\right)_0+
     2\int\! f({\bf p},{\bf p_1})\,{\partial n( {\bf p})
\over\partial  p_{1n}}\, dS_1dp_{1n}  \  ,
 \label{lanspa}
\eeq
where $dS$ is the surface element in momentum space. In obtaining
this relation, we have neglected a possible $\rho$ dependence of the
interaction function $f$; this simplification allows a
straightforward integration over coordinates.  Additionally,
we have employed the identity\cite{trio}
\beq
\left({\cal T}^k({\bf p})\right)_0=-\left({\partial G^{-1}({\bf
p})\over \partial {\bf p}}\right)_0= z\left({\partial
\epsilon({\bf p})\over \partial {\bf p}}\right)_0
\eeq
implied by the gauge invariance of the problem,
where $G$ is the single-particle Green function.

Assuming the ``bare'' spectrum $\epsilon^0({\bf p})$ to be stable,
potential points of divergence of the density of states are
associated---as in the preceding discussion of alternative scenarios
for the QCP in homogeneous matter---with (i) a vanishing of the
renormalization factor $z$ and consequent breakdown of the
quasiparticle picture {\it or} (ii) a change of sign of the
group velocity $v_n$, leading to a change of the connectivity of
the Fermi surface.

We restrict the analysis to the much-studied case of a
quasi-two-dimensional heavy-fermion metal with a quadratic lattice.
We also assume filling close to $1/2$.  Conventional belief would
hold that in this case, antiferromagnetic fluctuations play the key role
in the structure of the corresponding QCP.\cite{coleman1,steglichrev}
The relevant propagator takes the form
\beq
D({\bf k})\sim {1\over \xi^{-2}+({\bf k}-{\bf Q})^2}  \  ,
\label{anti}
\eeq
with the critical wave vector ${\bf Q}=(\pi/a,\pi/a)$, where $a$
is the lattice constant.  However, this viewpoint is insupportable. To
some extent, its weakness is already evident from the above
treatment of homogeneous matter; taking account of fluctuations
with $k_c>p_F\sqrt{2}$ results in a {\it suppression} of the
effective mass $M^*$.  Attending now to the anisotropic system,
based on the findings of Ref.~\onlinecite{cgy}, it may be
shown that in the postulated situation $z \to 0$, the first term
on the right-hand side of Eq.~(\ref{lanspa}) reduces to
$\partial \epsilon^0 ({\bf p})/ \partial  p_n$
to yield
\beq
\left({\partial \epsilon({\bf p})\over \partial p_n}\right)_0=
{\partial \epsilon^0 ({\bf p})\over
\partial  p_n}+g\int{\partial n({\bf p}_1)/ \partial  p_{1n}\over
\xi^{-2}+({\bf p}-{\bf p}_1-{\bf Q})^2} dS_1dp_{1n}  \  ,
\label{rela}
\eeq
where $g>0$.  In the case of
ferromagnetic or long-wave antiferromagnetic critical fluctuations,
the overwhelming contributions to the integral
term of Eq.~(\ref{rela}) come from a region ${\bf p}_1\simeq {\bf p}$
where the derivatives $\partial n({\bf p})/\partial p_n$ and
$\partial\epsilon^0 ({\bf p})/\partial p_n$ have opposite signs.
Consequently, accounting for these critical fluctuations
leads to a suppression of the group velocity $v_n$, promoting emergence
of the QCP.

What happens if instead one pursues the case of short-wave
antiferromagnetic
fluctuations, invoked in conventional scenarios of the QCP?
In this case, the so-called hot spots---points of the Fermi surface
connected by the antiferromagnetic vector ${\bf Q}$---contribute
predominantly to the integral term of Eq.~(\ref{rela}).
The quantities $\partial \epsilon^0({\bf p})/\partial p_n$
and $\partial n({\bf p}-{\bf Q})/\partial p_n$ now have the
{\it same} sign and thus cooperate to {\it enhance} rather than suppress
the value of the group velocity $v_n({\bf p})$.  We see then that
antiferromagnetic fluctuations actually have the effect of
{\it hampering} the emergence of the QCP in anisotropic systems, in
correspondence with the result obtained for homogeneous matter.
Accordingly, despite common belief, in heavy-fermion metals the
QCP cannot be attributed to antiferromagnetic fluctuations with
the critical wave vector ${\bf Q}=(\pi/a,\pi/a)$.

\subsection{Salient features of the topological  scenario for the QCP}

\begin{figure}[t]
\includegraphics[width=0.8\linewidth,height=0.5\linewidth]{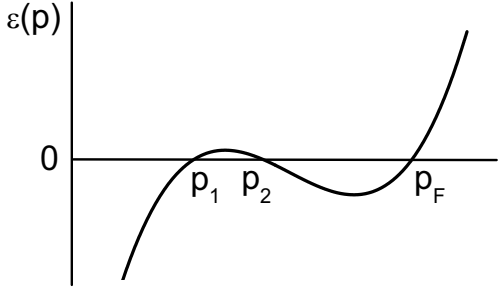}
\caption {Location of roots of the equation $\epsilon(p,T=0)=0$
measured from the chemical potential $\mu$ beyond the bifurcation
point} \label{fig:topo}
\end{figure}

In ordinary (``canonical'') Fermi liquids, there exists a one-to-one
correspondence between the single-particle spectrum $\epsilon(p)$
and the momentum $p$, at least in the vicinity of the Fermi
surface. This is equivalent to  asserting that the equation
\beq
{p^2\over 2M}+\Sigma(p,\varepsilon=0)=\mu  \
\label{bifm}
\eeq
has a single solution $p=p_F$, since the group velocity $v_F$ is
positive. Within FL theory, Eq.~(\ref{bifm}) reduces to the relation
\beq
\epsilon(p,T=0;n_F)=0 \  ,
\label{bif}
\eeq
where the energy $\epsilon(p)$, measured from the chemical
potential $\mu$, is evaluated with the Landau quasiparticle
momentum distribution $n_F(p)=\theta(p-p_F)$.  However, it
is a key ingredient of the topological scenario for the QCP that
at a critical value of some input parameter, specifically the
density $\rho_{\infty}$ or a coupling constant $g_{\infty}$,
the group velocity $v_F$ vanishes.  (This feature in demonstrated
in several numerical examples presented in Sec.~V.)

Beyond the critical point,
e.g.\ at $g>g_{\infty}$, Eq.~(\ref{bif}) acquires at least two
new roots (see Fig.~\ref{fig:topo}), triggering a {\it
topological} phase transition.\cite{volrev} Significantly, terms
proportional to $\epsilon\ln\epsilon$, which are present in the mass
operator $\Sigma$ of marginal Fermi liquids, do not enter
Eqs.~(\ref{bifm}) and (\ref{bif}).

\begin{figure}[t]
\includegraphics[width=0.95\linewidth,height=0.76\linewidth]{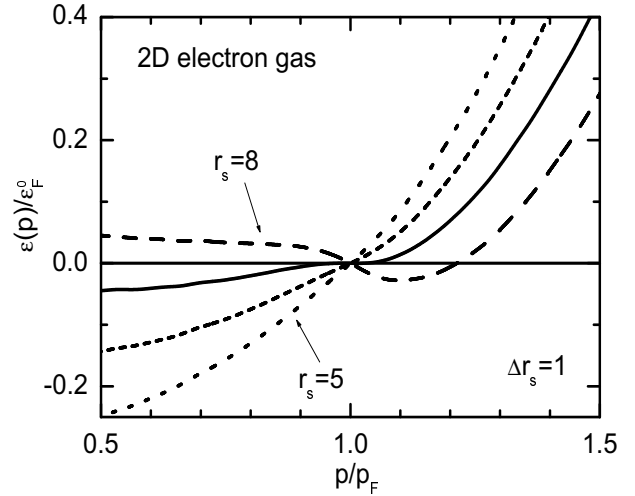}
\caption {Single-particle spectrum $\epsilon(p,T=0)$ of a
homogeneous 2D electron gas, measured from the chemical potential
$\mu$, in units of $\varepsilon_F^0=p^2_F/2M$, evaluated
microscopically \cite{bz} at $T=0$ for different values of the
dimensionless parameter $r_s=\sqrt{2}Me^2/p_F$.}
\label{fig:fc_2deg}
\end{figure}

The bifurcation point $p_b$ of these equations can emerge anywhere
in momentum space. Here we examine the case in which $p_b$
coincides with the Fermi momentum $p_F$, so that at the
critical density $\rho_{\infty}$, the group velocity
$v_F(\rho_{\infty})$ {\it changes its sign}.
This distinctive
feature of the {\it topological} scenario for the QCP is in
agreement with results of microscopic calculations of the
single-particle spectrum $\epsilon(p,T=0)$ of 2D electron
gas,\cite{bz} as illustrated in Fig.~\ref{fig:fc_2deg}. It is seen
that the sign of $v_F$ remains positive until the dimensionless
parameter $r_s$ attains a critical value $r_{\infty}\simeq 7.0$,
where $v_F$ vanishes.  At larger $r_s$, the group velocity
$v_F=\left(d\epsilon(p,n_F)/dp\right)_0$, evaluated with the
quasiparticle momentum distribution $n_F(p)$, is negative.
Having lost it stability, the Landau state is replaced by
a new state through the intervention of a topological
phase transition.

A qualitatively similar situation is to be expected in anisotropic
Fermi systems.  In support of this statement, we may refer to
the first example of the QCP, namely the Lifshitz saddle
point,\cite{lifshitz} which is characterized by a density of states
diverging at $T\to 0$ as $N(T,\rho_{\infty})\sim \ln (1/T)$.
This example has been elaborated specifically for anisotropic
2D electron systems in solids.  Close to the saddle point,
the QCP single-particle spectrum has the form $\epsilon({\bf
p},\rho_{\infty})=(1/2)(p^2_x/M_{xx}-p^2_y/M_{yy})$, with the
components $p_x$ and $p_y$ specifying the distance to the
saddle point in momentum space.  In the Lifshitz model, the
quantities $M_{xx}$ and $M_{yy}$ have comparable finite values.
If one of these parameters, say $M_{xx}$, becomes much larger
than $M_{yy}$, we arrive at the so-called extended saddle
point, which has been considered in connection with high-$T_c$
superconductivity in Ref.~\onlinecite{abr}.  The corresponding
phase transition, associated with the divergence of $M_{xx}$,
is also topological.

\section{Poincar\'e mapping for strongly correlated Fermi systems}

In this section we restrict considerations to homogeneous matter
at zero temperature.  Microscopic calculations of the
single-particle spectrum $\epsilon(p,T=0)$ are as yet available
only for several simple types of bare interactions between
particles. Moreover, there is no microscopic theory beyond the
point where the Landau state loses its stability. On the other
hand, given a phenomenological interaction function $f$,
Eq.~(\ref{lansp}) holds on both the sides of the topological QCP,
since the $z$-factor retains a nonzero value. Numerical iteration
is a standard approach to solution of an equation such as
Eq.~(\ref{lansp}). We shall see that the mathematical counterpart
of iteration, Poincar\'e mapping, which has been widely exploited
in theory of turbulence,\cite{lanh} is also instrumental in
elucidating the striking features inherent in solutions of
Eq.~(\ref{lansp}) beyond the QCP.

The discrete iterative map corresponding to Eq.~(\ref{lansp})
reads
\beq
{\partial\epsilon^{(j+1)}(p)\over\partial p} ={p\over M} +{1\over
3\pi^2} \int f_1( p, p_1){\partial n^{(j)}(p_1)\over\partial
p_1}p^2_1dp_1  \  ,
\label{lanit}
\eeq
the iterate $\mu^{(j+1)}$ for the chemical potential being
determined from the normalization condition (\ref{part}).
The index $j = 0,1,2, \ldots$ counts the iterations (zeroth, first,
second, $\ldots$).  The iterate $n^{(j+1)}(p)$ of the momentum
distribution $n(p)$ is generated by inserting the corresponding
spectral iterate $\epsilon^{(j+1)}(p)$ into Eq.~(\ref{dist}), which
at $T=0$ reduces to a Heaviside function $n(p)=\theta(-\epsilon(p))$.

\subsection{Poincar\'e mapping for systems with a nonsingular
interaction}

The key quantity of the Poincar\'e analysis in the Fermi liquid
problem is the group velocity
\beq
{d\epsilon^{(1)}(p)\over dp} ={p\over M}-{p^2_F\over
3\pi^2}f_1(p,p_F;\rho) \ ,
\label{eps1}
\eeq
evaluated by inserting the standard FL momentum distribution
$n_F(p)=\theta(p-p_F)$ into the right-hand side of
Eq.~(\ref{lanit}) as the zero iterate (starting
approximant) for the quasiparticle momentum distribution
$n(p)$.  In canonical Fermi liquids, for which the sign of
$v_F$ is positive and the function $\epsilon^{(1)}(p)$
has the single zero at $p=p_F$, the first iterate
$n^{(1)}(p)$ and all higher iterates for the distribution
$n(p)$ coincide with the $n_F(p)$,
this being a fixed point of the transformation.  However,
beyond the QCP the sign of the group velocity $v_F$
evaluated from Eq.~(\ref{eps1}) becomes negative,
and the first iterate $\epsilon^{(1)}(p)\equiv \epsilon(p;n_F)$
for the spectrum already has three zeroes
$p_1^{(1)}<p_2^{(1)}<p_3^{(1)}$, implying three kinks in
the momentum distribution.

At the second step, the evolution of the iteration process
follows different patterns, depending on the presence
or absence of a long-range component in the effective
interaction between quasiparticles (long range in
coordinate space).  We first consider the simpler
case in which the effective interaction $f(k)$, local
in coordinate space, has no singularities.  In this case
the interaction function $f_1(p,p_F)$ can then expanded
in a Taylor series in the variable $p-p_F$, and
Eq.~(\ref{lansp}) may be recast as a set of algebraic
equations.  Then, as the straightforward analysis demonstrates
and the numerical calculations described in Sect.~V confirm,
the shape
\beq
\epsilon(p,T=0)\sim (p-p_1)(p-p_2)(p-p_3)
\label{spbey}
\eeq
of the spectrum remains the same independently of the number of
iterations, with three roots $p_1<p_2<p_3$ specifying the location
of the three sheets of the Fermi surface and lying close
to the QCP Fermi momentum $p_{\infty}$.  Correspondingly,
the spectrum $\epsilon(p,T=0)$ changes smoothly in the momentum
regimes removed from the kinks, but oscillates in the interval
$[p_1,p_3]$.  The amplitude of the oscillation, i.e., the maximum
value
\beq
T_m=\max |\epsilon(p,T=0)|\  ,\quad  p_1<p<p_3 \,,
\label{tm}
\eeq
of the departure of $|\epsilon(p,T=0)|$ from 0, defines a new energy scale
of the problem: if temperature $T$ attains values comparable to $T_m$,
the kink structure associated with the multi-connected Fermi surface
is destroyed.

To evaluate $T_m$, we note that according to Eq.~(\ref{spbey})
the group velocity $d\epsilon(p)/dp$ is a parabolic function of
$p$, conveniently written as
\beq
{d\epsilon(p)\over dp}=-X+Y(p-p_m)^2  \  ,
\label{par}
\eeq
where $p_m$ determines the location of the minimum of the group
velocity.  Comparison of Eqs.~(\ref{spbey}) and (\ref{par}) leads
to the following set of equations,
\begin{eqnarray}
r_1+r_2+r_3&=&0  \  , \nonumber\\
r_1r_2+r_1r_3+r_2r_3&=& -{3X\over Y} \ ,
\nonumber\\
(p_m+r_1)^3-(p_m+r_2)^3+(p_m+r_3)^3&=&3\pi^2\rho \ ,
\label{root3}
\end{eqnarray}
for the three (shifted) roots $r_k=p_k-p_m$, with $k=1,2,3$.
The last of Eqs.~(\ref{root3}) is obtained from the normalization
condition (\ref{part}).

In the vicinity of the QCP where $X=X_{\infty}=0$, the parameter
$p_m$ and the positive quantities $X(\rho)$ and $Y(\rho)$ can be
shown to change linearly with $\rho-\rho_{\infty}$.  Simple
but lengthy algebra then yields
\beq
r_3\simeq -r_1\sim |\rho-\rho_{\infty}|^{1/2}\,, \quad
r_2\sim |\rho-\rho_{\infty}| \ .
\eeq
These results allow us to express relevant parameters in terms of
the difference $\rho-\rho_{\infty}$, namely (i) $\Delta p=p_3-p_1$,
which characterizes the range of the flattening of the spectrum
$\epsilon(p,T=0)$ beyond the QCP, (ii) the temperature $T_m$
associated with the crossover from standard FL behavior to NFL
behavior, and (iii) the zero-temperature density of states
$N(0)\sim \sum_k [d\epsilon(p=p_k)/dp]^{-1}$, which replaces
the ratio $M^*/M$ in the standard FL formulas for the specific
heat, spin susceptibility, etc.  We find
\beq
\Delta p\sim |\rho-\rho_{\infty}|^{1/2}, \;
T_m\sim |\rho-\rho_{\infty}|^{3/2}, \;
N(0)\sim |\rho-\rho_{\infty}|^{-1}  \  .
\label{relq}
\eeq
To our knowledge, the multi-connected Fermi surface, shown here
to arise in the homogeneous case,
was first considered in 1950 by Fr\"ohlich.\cite{frohlich}  Within
the Hartree-Fock (HF) approach, model variational calculations for
the ground-state energies of homogeneous systems leading to a
multi-connected Fermi surface were performed more than 20 years
ago,\cite{Vary} while the corresponding phase transition was
first discussed in terms of HF single-particle spectra in
Ref.~\onlinecite{baym} and later in
Ref.~\onlinecite{schofield}.  The original calculations of properties
of this transition on the basis of Eq.~(\ref{lansp}) were performed
in Refs.~\onlinecite{zb,shagp}. These calculations show that as
the coupling constant moves away from the QCP value $g_{\infty}$,
the number of sheets of the Fermi surface, which coincides with
the number of the roots of Eq.~(\ref{bif}), grows very rapidly, with
the distance between the sheets shrinking apace. Importantly, in all these
solutions, the relation
\beq
n^2(p)=n(p) \ ,
\label{imp}
\eeq
inherent in the standard FL picture, is still obeyed.

\subsection{2-cycles in Poincar\'e mapping for systems with a singular
interaction function $f$}

A remarkable feature of equations relevant to the turbulence problem
is the doubling of periods of motion, giving rise to dynamical
chaos.\cite{feig,lanh}  If now the nonlinear system corresponding
to Eq.~(\ref{lanit}) is considered within this context by associating
an iteration step with a step in time, one's first instinct is
to assert that such a phenomenon cannot occur when Poincar\'e
mapping is implemented, since chaos in the classical sense
cannot play a role in the ground states of Fermi liquids at $T=0$.
And indeed, this general assertion seems to be validated by the
results of the preceding subsection.  However, the Taylor expansion
of the interaction function $f_1(p,p_F)$ that we employed in
the above analysis fails in the case of long-range effective
interactions.  The Fourier transform $f(k\to 0)$ becomes
singular, and the previous analysis is inapplicable.

As an interesting example, we note that
such a singularity exists in the bare interaction
\beq
\Gamma^0({\bf p}_1,{\bf p}_2,{\bf k},\omega=0)
=-g {{\bf p}_1\cdot{\bf p}_2-({\bf p}_1\cdot{\bf k})
({\bf p}_2\cdot{\bf k})/k^2 \over k^2}
\label{cur}
\eeq
between quarks in dense quark-gluon plasma, wherein $g>0$.
\begin{figure}[t]
\includegraphics[width=1.0\linewidth,height=1.1\linewidth]{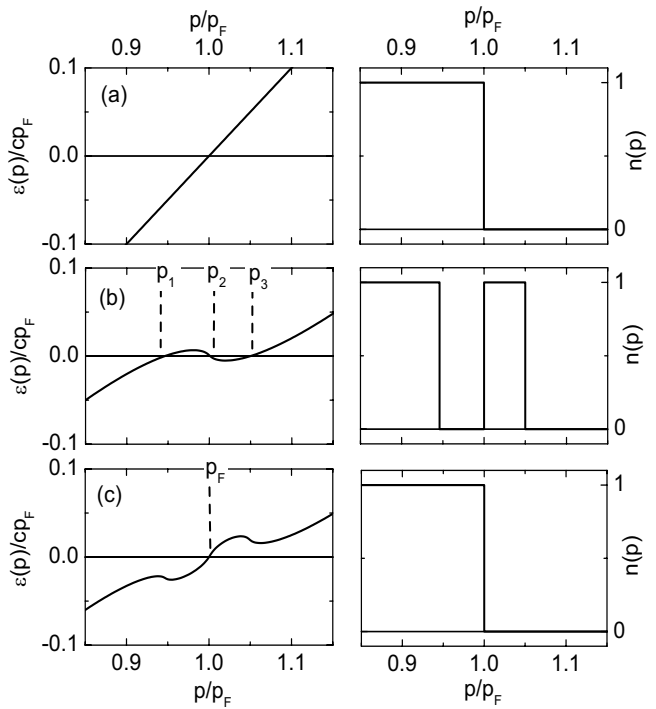}
\caption {Iterative maps for the quark-gluon plasma problem
(\ref{lqgp}) with the bare spectrum $\epsilon_p^0=cp$ and the
dimensionless parameter $\alpha=g/c=0.3$. Left panels: spectral
iterates $\epsilon^{(j)}(p)$ with $j=0,1,2$, in units of $cp_F$.
Right panels: momentum-distribution iterates $n^{(j)}(p)$.
The 2-cycle reveals itself in the coincidence between the first
and third right panels.} \label{fig:fc_cc_2c}
\end{figure}
In this example, Eq.~(\ref{lansp}) has the form\cite{baym}
\beq
{\partial\epsilon(p)\over \partial p} ={\partial\epsilon^0_p\over
\partial p} +g \int \ln {2p_F\over p-p_1}
{\partial n(p_1)\over \partial p_1} dp_1 \ ,
\label{lqgp}
\eeq
where $\epsilon^0_p\simeq cp$ is the bare single-particle
spectrum of light quarks.

The first iterate for the spectrum, evaluated from
Eq.~(\ref{lqgp}) with the distribution $n_F(p)$, has an infinite
{\it negative} derivative $d\epsilon^{(1)}/dp$ at the Fermi
surface.  It then follows that independently of the value of
the dimensionless coupling constant
$\alpha=g({\partial\epsilon^0_p/\partial p})^{-1}_0$,
Eq.~(\ref{bif}), with $x=p/p_F - 1$, takes the form
\beq
x\left(1-\alpha\ln {2\over x}\right)=0 \ ,
\eeq
and has three different roots $-x_0$, $0$, and $x_0$, with
$x_0=2\exp^{-1/\alpha}$. The corresponding first iterate of the
momentum distribution is $n^{(1)}(p)=\theta(x+x_0)-\theta(x)
+\theta(x-x_0)$.  The next iteration step yields $n^{(2)}(p)
\equiv n_F(p)$, so the standard FL structure of the momentum
distribution is recovered.  The nonlinear system enters a 2-cycle
that is repeated indefinitely; one is dealing with a 2-cycle
Poincar\'e mapping.

The first two iterations of the mapping process are illustrated in
Fig.~\ref{fig:fc_cc_2c}. The top-left panel of this figure shows
the bare spectrum $\epsilon^{(0)}(p)\equiv\epsilon^{0}(p)$.  The
first iterate of the spectrum, $\epsilon^{(1)}(p)$, appearing in
the middle-left panel, is evaluated by folding the kernel
$\ln(2p_F/(p-p_1))$ with the Fermi-step $n_F(p)$, shown in the
top-right panel. The spectrum $\epsilon^{(1)}(p)$ possesses three
zeroes: $p_1<p_F$, $p_2=p_F$, and $p_3>p_F$, implying that the
first iterate $n^{(1)}(p)$, drawn in middle-right panel, describes
a Fermi surface with three sheets.  This distribution differs from
the ordinary Fermi step {\it only} in the momentum interval
$-x_0<x<x_0$.  The next iterate, $\epsilon^{(2)}(p)$ (bottom-left
panel), again has a single zero $p_F$, and the corresponding
momentum distribution $n^{(2)}(p)$ (bottom-right panel) coincides
identically with $n_F(p)$.

\begin{figure}[t]
\includegraphics[width=1.0\linewidth,height=1.1\linewidth]{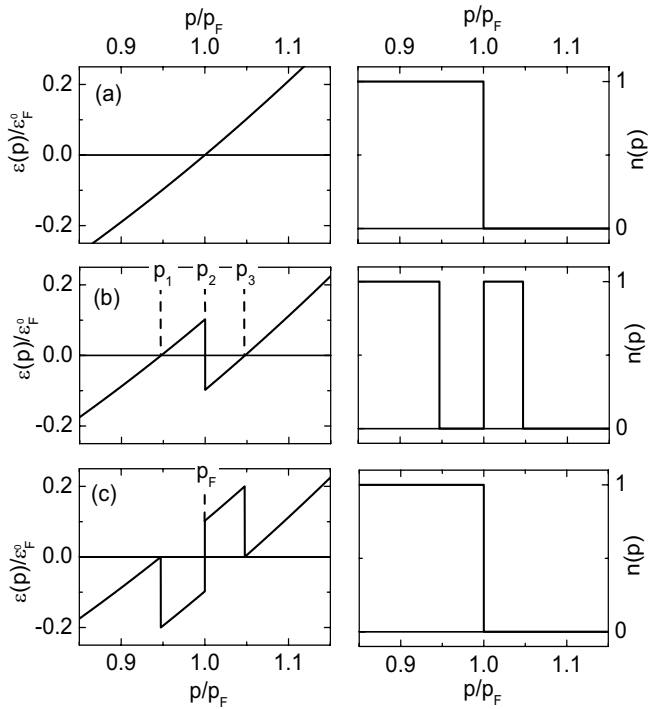}
\caption {Same as in Fig.~\ref{fig:fc_cc_2c} for the discretized
Nozi\`eres model (\ref{nozi}) with $g=0.2\varepsilon_F^0$.}
\label{fig:fc_n_2c}
\end{figure}

A 2-cycle Poincar\'e mapping also arises in treating the
Nozi\`eres model,\cite{noz} for which the interaction function
$f$ has the limited singular form $f({\bf q})=g\delta({\bf q})$
with $g>0$.  In this model, the iterative maps (illustrated in
Fig.~\ref{fig:fc_n_2c}) are generated from the equation
\beq
\epsilon^{(j+1)}(p)+\mu^{(j+1)}=p^2/2M+gn^{(j)}(p)\ ,
\label{nozi}
\eeq
along with the normalization condition (\ref{part}) for
$n^{(j+1)}(p)=\theta(-\epsilon^{(j+1)}(p))$.  Here, the odd
iterates $n^{(2j+1)}(p)$ of the momentum distribution
deviate from the $n_F(p)$ in the interval
$-g/4\varepsilon^0_F<x<g/4\varepsilon^0_F$, but in
even iterations, the Fermi step reappears intact.

Numerical analysis demonstrates that similar 2-cycles arise
when Poincar\'e mapping based on Eq.~(\ref{lanit})
is implemented for other systems having an interaction
function that is singular at $k\to 0$ (see Sec.~V).
In all these cases, the emergence of a 2-cycle turns out to
be an unambiguous signal of the instability of the standard
Landau state.  Moreover, the domain of momentum involved in
the cyclic behavior is almost identical with the domain within
which an improved iteration algorithm fails to converge
(see Sec.~III.D).  As will be seen, this concurrence is
significant in that the associated volume of momentum
space provides a measure of the entropy stored in the
exceptional ground state that replaces the Landau state.

\subsection{2-cycles in  Poincar\'e mapping for finite systems}

In finite Fermi systems---nuclei, atoms, atomic clusters,
quantum dots, molecules, $\ldots$---integration over momenta
in Eq.~(\ref{lanit}) is replaced
by summation over single-particle quantum numbers.  As a concrete
illustration, let us explore a schematic model\cite{twolevel} of a
spherical nucleus, in which the single-(quasi)particle energies are
independent of the magnetic quantum number $m$.  We consider
two neutron levels available for filling in an open shell, denoted
$0$ and $+$.

In the presence of a quasiparticle interaction $f$,
the energies of the two levels are influenced when neutrons are
added to the system.  With this in mind, the energies of levels $0$
and $+$ when $N$ neutrons are added to level $0$ and none added
to level $+$ are denoted $\epsilon_0(N,0)\equiv \epsilon_0(N)$
and $\epsilon_+(N,0)\equiv \epsilon_+(N)$, respectively.  The initial
distance between the two levels, with no added neutrons, is
$D(0,0) = \epsilon_+(0,0) - \epsilon_0(0,0)= D_0 > 0$, having
designated $0$ as the lower of the two levels.  The behavior of
the level distance $D(N,0)=\epsilon_+(N,0)-\epsilon_0(N,0)$
as neutrons are added is crucial to the behavior of the system.
To facilitate the argument, we adopt a highly simplified interaction
between neutrons, retaining in $f$ only a principal\cite{migdal,ring}
$\delta({\bf r})$-like component. Also for simplicity, the relevant matrix
elements for determining the energy shifts are reduced to two, namely
$f_{00}=f_{++}$ and $f_{0+}=f_{+0}$, which (with $k=0,+$) are calculated
as
\beq
f_{kk'}=\int R_k^2(r)
f\left(\rho(r)\right)R_{k'}^2(r){r^2dr\over 4\pi}
\label{mel}
\eeq
in terms of the radial parts of the corresponding single-particle
wave functions.  We then have
\begin{equation}
\epsilon_0(N)=\epsilon_0+Nf_{00}, \quad
\epsilon_+(N)=\epsilon_0+D_0+Nf_{+0} \  , \label{mig}
\end{equation}
so that the distance $D(N,0)$ varies as
\beq
D(N,0)=D_0+N\left( f_{+0}-f_{00}\right) \  .
\label{diff}
\eeq

Significantly, in both the atomic and nuclear problems, the sign
of the difference $f_{0+}-f_{00}$ is {\it negative},\cite{twolevel}
so that the function $D(N,0)$ falls off as the number $N$
of added neutrons increases.  The standard FL scenario, in which
all added quasiparticles occupy the level $0$, remains valid as
long as the level distance $D(N,0)$ remains positive.  As seen
from Eq.~(\ref{diff}), this distance changes sign when $N$
reaches the critical number $N_c=D_0/(f_{00}-f_{+0})$.  Thus,
at $N>N_c$,
\beq
D(N,0)=(N-N_c)\left( f_{+0}-f_{00}\right)<0 \  ,
\label{ineq1}
\eeq
forcing all the quasiparticles to resettle into the level $+$.

To perform the next iteration, one calculates the
distance $D(0,N)\equiv \epsilon_0(0,N)-\epsilon_+(0,N)$
between levels $0$ and $+$ for the case that all the quasiparticles
occupy level $+$, obtaining
\beq
D(0,N)=D_0+N \left(
f_{+0}-f_{++}\right)\simeq-(N-N_c)\left(f_{00}- f_{+0}\right) \  .
\eeq
Since $N$ exceeds the critical number $N_c$, the sign of this
quantity is {\it negative}.  This implies that the priority for
filling reverses again, requiring that all the quasiparticles
return to the level $0$.  This 2-cycle is repeated {\it indefinitely}.

In the weak coupling limit where $f_{00}\to 0$, the critical
number $N_c$ diverges, and filling of the level $0$ is completed
before $N$ reaches $N_c$.  In this case, level-filling occurs
normally, the 2-cycle being unattainable.  However, in the atomic
problem, the magnitude of the matrix elements of the Coulomb
interaction between the orbiting electrons significantly exceeds
the distance between neighboring single-particle levels.  As a
consequence, 2-cycles emerge at $N\simeq 1$ in the iterative
maps for atoms of almost all elements not belonging to
the principal groups.  The implication is that the electronic
systems of these elements {\it do not obey standard FL theory}.
The same conclusion is valid for many heavy atomic nuclei
with open shells.

Naturally, the occurrence of such 2-cycles in the analysis
of level-filling in finite Fermi systems presents a dilemma
that must be resolved.  Here (and in the other examples) such
cyclic behavior is obviously unphysical.  Here the resolution
lies in the merging of single-particle levels and their
{\it partial occupation}, hence in behavior that conflicts
with standard FL theory yet in fact maintains consistency
within the broader framework of Landau theory.\cite{twolevel}

\subsection{Modified Poincar\'e mapping and new insight from
chaos theory}

Apparently, the occurrence of persistent 2-cycles in the
iterative maps of Eq.~(\ref{lanit}) prevents us from finding
true solutions of the fundamental Landau equation (\ref{lansp})
beyond the QCP for a specific class of Fermi systems possessing
singular effective interactions.  It can be argued, however,
that this failure is a consequence of the inadequacy of the
iterative procedure employed, which works perfectly on the
FL side of the QCP.  Indeed, a refined procedure that mixes
iterations does allow one to avoid the emergence of these 2-cycles.
Nevertheless, the improved procedure possesses the same feature:
the iterations do not converge, although the pattern of their
evolution becomes more complicated and---as will be seen---both
intriguing and suggestive.

\begin{figure}[t]
\includegraphics[width=1.0\linewidth,height=1.41\linewidth]{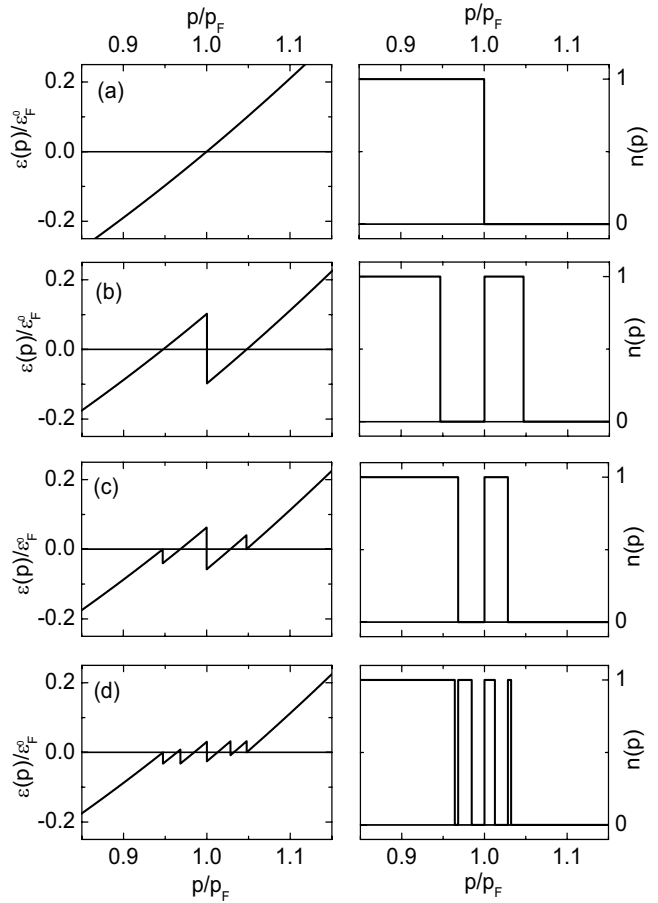}
\caption {Iterative maps for the discretized Nozi\`eres model (\ref{nozi1})
when inputs taken from the two preceding iterations are mixed with
the parameter $\zeta=0.2$.  Left panels: spectra $\epsilon^{(j)}(p)$
with $j=0,1,2,3$, in units of $\varepsilon^0_F$. Right panels:
momentum distributions $n^{(j)}(p)$.
} \label{fig:fc_n_mix}
\end{figure}

By way of illustration let us consider a modified Poincar\'e
mapping for the Nozi\`eres model, again with coupling constant
$g=0.2\varepsilon^0_F$.  Choosing a mixing parameter $\zeta$,
the equation
\begin{eqnarray}
 \lefteqn{\epsilon^{(j+1)}(p)+\mu^{(j+1)}=
 (1-\zeta)\,[\epsilon^{(j)}(p) +\mu^{(j)}]} \qquad\qquad
 \nonumber \\
 & &+\zeta\,[p^2/2M+gn^{(j)}(p)] \qquad\qquad
\label{nozi1}
\end{eqnarray}
is used to generate the iterative maps shown in Fig.~\ref{fig:fc_n_mix}.
Recovery of the ordinary Fermi distribution $n_F(p)$,
an inherent feature of the standard iteration procedure at
the second iteration (see Fig.~\ref{fig:fc_n_2c}),
no longer occurs; indeed, the sequence of iterations fails
to converge, even to a limit cycle.  The number of sheets
remains three at the second iteration.  At the third,
however, seven sheets of the Fermi surface emerge, and
the number of sheets continues to increase in successive
iterations.  At the same time, the distance between
the sheets continues to narrow, since the domain of momentum
space in which the improved iteration procedure does not
converge remains almost the same as that in which the standard
procedure finds a 2-cycle.  This phenomenon of proliferating
Fermi sheets recapitulates a scenario envisioned in the seminal
study by Pethick et al.\cite{baym} of the quark-gluon plasma
based on the interaction function (\ref{cur}).

Treating the number $j=1,2,3,\ldots$ of the iteration as a discrete
time step, the sequence of the pictures in the left column of
Fig.~\ref{fig:fc_n_mix} shows the ``temporal'' evolution of
the quasiparticle spectrum.  At any time-step $t$, the single-particle
energy $\epsilon(p,t)$ falls off steadily as $p$ goes to zero, while
its sign changes unpredictably in a finite region $\Omega_t$ of momentum
space adjacent to the Fermi surface.  These erratic changes of sign induce
unpredictable jumps of the occupation numbers $n(p,t)$ between the
two values 0 and 1.  Significantly, at $T=0$ and as $t$ goes to
infinity, the region of momentum space in which iterations of
Eq.~(\ref{lansp}) fail to converge tends to a definite
limit $\Omega$.  The lack of convergence of the iteration process
may be attributed to the presence of a kind of ``quantum chaos.''
Since entropy is a natural byproduct of chaos, we tentatively
identify $2\Omega\ln 2$ as its measure in the present context,
where $\Omega$ now denotes the numerical volume, in momentum
space, of the domain of non-convergence, and the factor 2 comes
from the two spin projections.
Thus, $2\Omega\ln 2$ is interpreted as a special entropy associated
with a Fermi system for which iteration of Eq.~(\ref{lansp})
does not converge to a solution.

This naive formula for the special entropy can be refined.  To do
so, we introduce a ``time-averaged'' single-particle energy
${\overline{\epsilon}}(p)$ by appropriating a standard formula from
statistical physics,
\beq
{\overline{\epsilon}}(p)=\lim\limits_{T\to \infty}
{1\over T}\int\limits_0^T\epsilon(p,t) dt\equiv
\lim\limits_{N\to\infty} {1\over N}\sum\limits_0^N\epsilon^{(j)}(p) \  ,
\label{epsi}
\eeq
in which the iterates for $\epsilon(p)$ are averaged over the
fictitious time $t$.  In the same way, we introduce a
corresponding time average ${\overline{n}}(p)$ of the iterates of
the momentum distribution $n(p)$.  The relation between the
two time averages, simply
\beq
{\overline{\epsilon}}(p)=p^2/2M -\mu +g {\overline{n}}(p)  \  ,
\eeq
stems from Eq.~(\ref{nozi}).
Remarkably, the mixing parameter $\zeta$ cancels out in arriving
at this relation.

Obviously, ${\overline{n}}(p)$ only takes values 0 or 1 wherever the
iterations converge, and then ${\overline{\epsilon}}(p)$ is a parabolic
function of $p$ coinciding with the true single-particle energy.
However, as seen from  Eq.~(\ref{epsi}) and verified by results
shown in the left column of Fig.~\ref{fig:fc_n_mix}, the
function ${\overline{\epsilon}}(p)$ {\it vanishes identically} in
the domain $\Omega$ where iterations fail to converge, yielding
\beq
n_*(p)= {\mu -p^2/2M\over g}\  , \quad {\bf p}\in \Omega \  .
\label{nav}
\eeq
The notation $n_*(p)$ has been introduced to signify a smoothed
momentum distribution, determined by averaging iterates
$\epsilon(p,t)$ for the single-particle energy
according to the prescription (\ref{epsi}).

Thus, by implementing (i) a modified Poincar\'e mapping
procedure based on Eq.~(\ref{nozi1}) together with (ii)
``time-averaging'' of iterates for $\epsilon(p)$ and $n(p)$
in the manner of Eq.~(\ref{epsi}), we have found a solution of
the problem that satisfies the Pauli principle in the domain
$\Omega$ where the sequence of iterates does not converge.
The solution so obtained is independent of the parameters
specifying the refined iteration procedure.
The boundaries $p_i$ and $p_f$ of the momentum interval
$p_i<p_F<p_f$ defining the domain of non-convergence are
determined by the conditions $n_*(p_i)=1$ and $n_*(p_f)=0$.
Thus, the quasiparticle momentum distribution $n(p)$ corresponding
to the new solution, hereafter written as $n_*(p)$, is given by
1 and 0, respectively, at $p \leq p_i$ and $p \geq p_f$,
and by Eq.~(\ref{nav}) in between.

Now we are prepared to introduce a Kolmogorov-like entropy
\beq
S_*=-2\int [n_*(p)\ln n_*(p) +(1-n_*(p))\ln (1- n_*(p))] d\upsilon  \ ,
\label{kolm}
\eeq
in which the second term of the integrand accounts for the equal
status of particles and holes.
According to the definition of $n_*(p)$, the integrand in Eq.~(\ref{kolm})
vanishes outside the domain $\Omega$, and we see that $S_*$ is essentially
proportional to the volume of this domain (also called $\Omega$).
Setting $n_*(p)$ inside $\Omega$ equal to a typical value 1/2, one does
indeed arrive at $S_*=2\Omega \ln2$.

This program is to be implemented similarly when dealing with
other Fermi systems for which iteration of Eq.~(\ref{lansp})
does not converge to a solution.  The first step is to integrate
Eq.~(\ref{lansp}) and find an explicit relation connecting the
spectrum $\epsilon(p)$ with the momentum distribution $n(p)$.
The average quantity ${\overline{\epsilon}}(p)$ is then constructed
by means of Eq.~(\ref{epsi}).  Since this quantity vanishes
identically in the domain $\Omega$ where iteration fails
to converge, we obtain a closed equation
\beq
{\overline{\epsilon}}(p,n_*)=0   \ ,  \quad {\bf p} \in \Omega \,,
\label {topi}
\eeq
for finding the smoothed, NFL component of the momentum
distribution $n_*(p)$ and calculating the entropy $S_*$.

The {\it possibility} of a nonzero value of the entropy at $T=0$
beyond the QCP is an inherent feature of the topological scenario
being explored within the original Landau framework. The
{\it existence} of such an entropy excess for a system whose
interaction function $f$ has a long-range component means that
the associated ground state is statistically degenerate.\cite{lanl}
These findings offer a new perspective not only on the QCP itself,
but also on the concept of ``quantum chaos''---suggesting a
novel role for the phenomenon of chaos in quantum many-body
systems.  The underlying connections and their implications
will be developed in depth in a separate paper.

\section{Fermion condensation as a species of topological phase transition}
An equation identical to (\ref{topi}) can be derived in another
way,\cite{ks} based on the premise\cite{lan} that the ground-state
energy $E$ of the system is a functional of the quasiparticle
momentum distribution $n(p)$.  For canonical Fermi liquids, the
minimum of $E[n]$ is attained at the boundary point
$n_F(p)=\theta(p-p_F)$ of the manifold $\{n\}$ containing all
candidates for the function $n(p)$ that satisfy the Pauli
restriction $0 \leq n \leq 1$ and particle-number conservation. In
Ref.~\onlinecite{ks}, the minimum of the model functional
\beq
E=\int {p^2 \over 2M} n({\bf p})\, d\upsilon +{f\over 2}\int
   {n({\bf p}_1)n({\bf p}_2)\over |{\bf p}_1-{\bf p}_2|} \,
   d\upsilon_1\,d\upsilon_2
\label{ksmodel}
\eeq
was found.  The Fermi step $n_F(p)$ turns out to be the true
ground-state quasiparticle momentum distribution {\it only}
if correlations are rather weak.  Otherwise a topological phase
transition occurs at some critical coupling constant $f_c$.
At $f>f_c$, a new distribution is determined from the variational
condition\cite{ks}
\beq
{\delta E\over \delta n({\bf p})}= \mu \  ,   \quad {\bf p} \in \Omega \  .
\label{var}
\eeq
For the above model functional, this condition takes the explicit
form
\beq
{p^2\over 2M}+f\int
{n({\bf p}_1)\over |{\bf p}_1-{\bf p}|} d\upsilon_1
= \mu  \  ,\quad {\bf p} \in \Omega \  .
\label{ele}
\eeq
This equation is easily solved for $n(p)$ by exploiting a
transparent analogy with a system of charged particles moving in
an external elastic field.  The result, yielding $n_*(p) = {\rm
const.} < 1$ at $p < p_f$ and $n_*(p)=0$ otherwise, is drastically
different from the familiar FL solution.  Importantly, the {\it
same} result is obtained when the iteration/averaging procedure
established in Sec.~III.D is applied to Eq.~(\ref{ele}). For the
Nozi\`eres model\cite{noz} studied in Subsec.~III.D,
Eq.~(\ref{var}) may also be solved analytically, and the solution
is again in agreement with the result (\ref{nav}) of the procedure
introduced there.

Prior to discussing this coincidence in more detail, let us focus on
Eq.~(\ref{var}) itself.  This variational condition, studied rather
extensively during the last decade,\cite{ks,vol,noz,physrep,khv,schuck,normfc,kzphysb,ikk,kzc2005,yak,shagrev} is generic.
To illuminate its nature and conceptual status, we invoke a
mathematical correspondence of the functional $E[n(p)]$ with
the energy functional $E[\rho(r)]$ of statistical physics. If the
interactions are weak, the latter functional attains its minimum
value at a density $\rho$ determined by the size of the
vessel that contains it, and which it fills uniformly.  In
such cases, solutions of the variational problem evidently
describe {\it gases}.  On the other hand, if the interactions
between the particles are sufficiently strong, there arise
nontrivial solutions of the variational condition
\beq
{\delta E[\rho]\over \delta \rho( r)}=\mu
\label{varrho}
\eeq
that describe {\it liquids}, whose density is practically
independent of boundary conditions.

The energy functional $E[n]$ of our quantum many-body problem must
have two analogous types of solutions, with an essential
difference: solutions $n_*(p)$ of the variational condition
(\ref{var}) must satisfy the condition $0\leq  n(p)\leq 1$. This
condition cannot be met in weakly correlated Fermi systems, but it
can be satisfied in systems with sufficiently strong correlations.

Turning now to the coincidence between solutions of Eq.~(\ref{var})
and Eq.~(\ref{topi}), we note that the quasiparticle energy is
by definition just the derivative of the ground state $E$ with
respect to the quasiparticle momentum distribution $n(p)$, or,
referred to the chemical potential, just
$\epsilon(p)=\delta E/\delta n(p)-\mu$.  Consequently, the
variational condition (\ref{var}) may recast as an equation
\beq
\epsilon(p,n_*)=0  \ , \qquad {\bf p} \in \Omega  \,,
\label{bifc}
\eeq
that is in fact identical to Eq.~(\ref{topi}).  This equivalence allows
us to use the same notation $n_*$ for the NFL momentum distribution
$n(p)$ in the two formulations of the problem.

The above analysis and discussion make it clear that the most
salient of the unorthodox features we have uncovered in applying the
original Landau quasiparticle formalism to the behavior of strongly
correlated Fermi systems beyond the QCP is that at $T=0$ the
single-particle spectrum $\epsilon(p)$ becomes {\it completely flat}
over a {\it finite domain} $\Omega$ of momenta adjacent to the Fermi
surface.  One may envision this phenomenon as a swelling of the
Fermi surface.  Another generic feature, concomitant with
flattening of the single-particle spectrum, is {\it partial
occupation of single-particle states of given spin}, i.e.,
$n(p)$ is no longer restricted to the values 0 and 1, but may
take any value in the interval $[0,1]$.

This extraordinary behavior is associated with a topological
phase transition fundamentally different from the one which features
a proliferating number of sheets of the Fermi surface, in that
the roots of the equation $\epsilon(p)=0$ now form a continuum
instead of a countable set.  Since the Landau quasiparticles
experience no damping, the single-quasiparticle
Green function becomes
\begin{equation}
G(p,\varepsilon)= {1-n_*( p)\over \varepsilon{+}i\delta}+ {n_*( p)
\over \varepsilon{-}i\delta} \,, \quad  \  {\bf p} \in \Omega \,,
\label{green}
\end{equation}
in the domain $\Omega$ of vanishing $\epsilon(p,n)$ and
retains its FL form for $p$ outside $\Omega$.
This particular structure of the Green function (\ref{green})
may be characterized by a {\it topological charge} defined as
the integral
\cite{vol,volrev}
\beq
{\cal N}=\int\limits_{\gamma}  G(p,\xi)\, \partial_l G(p,\xi)
{dl\over 2\pi i}  \  ,
\eeq
where the Green function is considered on the imaginary axis of
energy $\varepsilon=i\xi$ and the integration is performed over a
contour in $({\bf p},\xi)$-space that embraces the Fermi surface.
For canonical Fermi liquids and systems with a multi-connected
Fermi surface, the topological charge ${\cal N}$ is an integer,
whereas for the more exotic states characterized by a completely
flat portion of the spectrum $\epsilon(p)$, its value is
{\it half-odd-integral}.\cite{vol,volrev}

Aided by the variational condition (\ref{var}), one can
elucidate what happens when a quasiparticle with momentum ${\bf p}\in
\Omega$ is added to the system, assumed again to be homogeneous.
In contrast to what happens with a canonical Fermi liquid,
the addition of just one quasiparticle now induces a rearrangement of
the {\it whole} distribution function $n_*(p)$.  This implies
that the kind of system being considered cannot be treated
as a gas of interacting quasiparticles, even though the original
Landau quasiparticle concept still applies.

The set of states for which Eq.~(\ref{bifc}) [or Eq.(\ref{bifn})]
is satisfied has been called the {\it fermion condensate}
(FC),\cite{ks} while the topological phase transition in which
the Fermi surface swells from a line to a surface in 2D, or from
a surface to a volume in 3D, is otherwise known as {\it fermion
condensation}.  In finite systems the phenomenon of fermion
condensation exhibits itself as {\it merging} of neighboring
single-particle levels.\cite{twolevel}  Unfortunately, the terms
fermion condensation and fermion condensate have unnecessarily
promoted controversy.  Theorists are condition to think that in
contrast to bosons, fermions cannot condense, because fermions
cannot occupy the same quantum state. However, in
everyday life, ``condensation'' means simply a {\it dramatic
increase of density}.  For example, people use the word
to describe what occurs when the vapor in clouds forms
into liquid drops that fall as rain.  On the other hand,
statistical physics deals with occupation numbers rather than
with wave functions, and the thermodynamic properties of a Bose gas at
low temperatures are properly evaluated by treating Bose condensation
as a process in which a macroscopic number of bosons
occupy the same zero-momentum single particle state, all with
single-particle energy equal to the chemical potential $\mu$.

Ideally, in experimental
measurements of neutron scattering on liquid $^4$He, Bose
condensation is reflected in a sharp peak in the density of
states $\rho(\epsilon)=\rho_c\delta (\epsilon)$, with a
prefactor $\rho_c$ that is to be interpreted as the density
of the Bose condensate. In strongly correlated Fermi systems,
fermions are also capable of condensation in much the same
sense: a macroscopic number can have the same energy
$\mu$.  If so, it follows from Eq.~(\ref{green}) that the
zero-temperature density of states, which is associated
with the integral of the imaginary part of the retarded
Green function $G_R(p,\epsilon)$ over momentum space,
has the same kind of peak $\rho(\epsilon)= \rho_c\delta(\epsilon)$,
 where $\rho_c$ is the now the FC density.  We hasten to
add that although a macroscopic number of fermions have
the same energy, they possess different momenta; hence
the existence of states containing a FC does not contradict
the Pauli exclusion principle.

In spite of the similarities between boson and fermion condensation,
an important difference between the structure functions describing
them must be noted.  In the Bose case, a macroscopic number of
particles reside in condensate for all temperatures lower than
the critical temperature, with all the bosons having zero momentum
and energy $\epsilon(p)=0$.  On the other hand, in the Fermi case,
any elevation of the temperature from zero acts to lift the
degeneracy of the single-particle spectrum in the domain $\Omega$.
Indeed, since a minute temperature elevation does not affect the FC
momentum distribution, Eq.~(\ref{dist}) can be inverted to
yield \cite{noz}
\begin{equation}
\epsilon(p,T\to 0)= T\ln {1-n_*( p)\over n_*(p)} \ , \qquad {\bf p} \in
\Omega  \ .
\label{spet}
\end{equation}
Thus, the dispersion of the single-particle spectrum of systems
having a FC turns out to be {\it proportional to temperature}, in
contrast to the situation for canonical Fermi liquids,
where it is independent of $T$.  This distinction can be
employed for unambiguous delineation of the boundaries of
the domain $\Omega$ occupied by the FC at finite $T$.

In nonhomogeneous systems, finite or infinitely extended, appropriate
single-particle states are no longer eigenstates of momentum
${\bf p}$, so Eq.~(\ref{bifc}) must be replaced by
\beq
\epsilon_{\lambda}=0\  , \quad \lambda\in \Omega  \  ,
\label{bifn}
\eeq
where $\lambda$ is set of quantum numbers specifying
the single-particle state.  At finite temperatures, the FC degeneracy
is lifted according to the same formula (\ref{spet}) as for
the homogeneous case.

\section{Numerical calculations }

\begin{figure}[t]
\includegraphics[width=1.0\linewidth,height=0.72\linewidth]{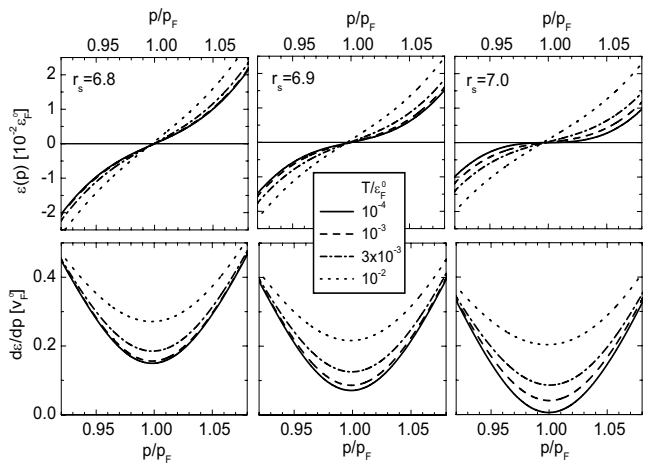}
\caption {Single-particle spectrum $\epsilon(p)$ (top panels) and
derivative $d\epsilon(p)/dp$ in units of $v_F^0=p_F/M$ (bottom
panels), as calculated for the interaction model (\ref{model_2pf})
with parameters $g_2=0.16$ and $\beta_2=0.14$.  These quantities
are plotted as functions of $p/p_F$ at four (line-coded) temperatures
expressed in units of $10^{-2}\,\varepsilon_F^0$ and at $r_s=6.8$
(left column), $r_s=6.9$ (middle column), and $r_s=7.0$ (right
column).} \label{fig:eg_fl}
\end{figure}

In this section, we discuss the results of numerical calculations
of the single-particle spectra $\epsilon(p,T)$ and quasiparticle
momentum distributions $n(p,T)$ beyond the QCP.  We compare the
temperature evolution of these quantities in a system for which
the interaction function $f(k)$ is singular at $k \to 0$, with
their temperature dependence in systems for which $f(k)$ has no
singularities.

We consider model quasiparticle interaction functions $f(k)$ that
depend only on the difference ${\bf k}={\bf p}-{\bf p}_1$. In this
case one can integrate the relation (\ref{lansp}) over momentum
and arrive at the integral equation
\beq
\epsilon(p) = {p^2\over M} - \mu + 2\int\! f({\bf p}-{\bf p_1})\,
         n(\epsilon(p_1))\, d\upsilon_1
\label{lansp1}
\eeq
for the quasiparticle spectrum $\epsilon(p)$, the chemical
potential $\mu$ being determined by the normalization condition
(\ref{part}). To evaluate the $T$-dependent spectrum $\epsilon(p,T)$
entering all thermodynamic and transport integrals, we substitute
the formula (\ref{dist}) into the right-hand side of Eq.~(\ref{lansp1}).
We employ an iteration procedure that mixes iterations for numerical
solution of Eq.~(\ref{lansp1}), with a mixing parameter $\zeta$ having
the same meaning as in Eq.~(\ref{nozi1}). It is worth noting that
although at low temperature the momentum distribution $n(p,T)$ is
smoothed somewhat, one meets the same difficulties with iterative solution
of Eq.~(\ref{lansp1}) for singular interaction functions as in the
case of $T=0$ when the mixing parameter $\zeta$ is much larger
than the ratio $T/\varepsilon^0_F$. However as soon as $\zeta$
becomes comparable with this ratio, stable convergence of the
iteration procedure is achieved. In the case of non-singular interactions,
the requirements to be met by $\zeta$ are less stringent; in particular,
$\zeta$ can be taken as large as some tens of $T/\varepsilon^0_F$.
Extremely low temperatures are not accessible to the numerical analysis
since the mixing parameter $\zeta$ should then be taken so small that
the CPU time required to attain reasonable accuracy is unreasonably large.

The numerical calculations were performed on a momentum grid with
the step size $10^{-4}p_F$. The mixing parameter
$\zeta=10^{-4}\div 10^{-2}$ was taken to achieve stable
convergence of iteration procedure. The accuracy of the numerical
solution, measured by the maximum discrepancy between the l.h.s.\
of Eq.~(\ref{lansp1}) and its r.h.s., was fixed at
$10^{-7}\varepsilon_F^0$. The number of iterations necessary to
reach this accuracy was about $10^2\div 10^4$, depending on
$\zeta$. Convergence of the iteration procedure is a warranty that
an obtained solution provides a local minimum of the free energy
functional $F[n]=E[n]-S[n]T-\mu N$.

We first describe the situation for regular interactions.
If the analysis carried out in Sec.~III is correct, then at $T=0$
the condition (\ref{imp}) stays in effect, but the system possesses a
multi-connected Fermi surface.  At low $T$ below the new temperature
scale $T_m$, the thermodynamic properties of the system still
follow standard Landau FL theory.
However, the enhanced density of states, whose value we have
determined to be inversely proportional to the difference
$|\rho-\rho_{\infty}|$, produces a great enhancement of key
thermodynamic characteristics, notably the spin susceptibility
$\chi(T\to 0)$ and the ratio $C(T)/T$ at $T \to 0$.  If the
temperature $T$ reaches values comparable to $T_m$, the
sharp kink structure seen in the momentum distribution
$n(p,0)$ associated with the multi-connected Fermi surface
becomes smeared, and at $T>T_m$ the function $n(p,T)$
becomes continuous and almost independent of $T$ in the momentum
interval $[p_1,p_3]$ between the sheets.  According to
Eq.~(\ref{spet}), the dispersion of the single-particle
spectrum $\epsilon(p,T)$ then becomes proportional to $T$,
as in systems with a FC.  All these features are confirmed
in the calculations.

\begin{figure}[t]
\includegraphics[width=1.0\linewidth,height=1.0\linewidth]{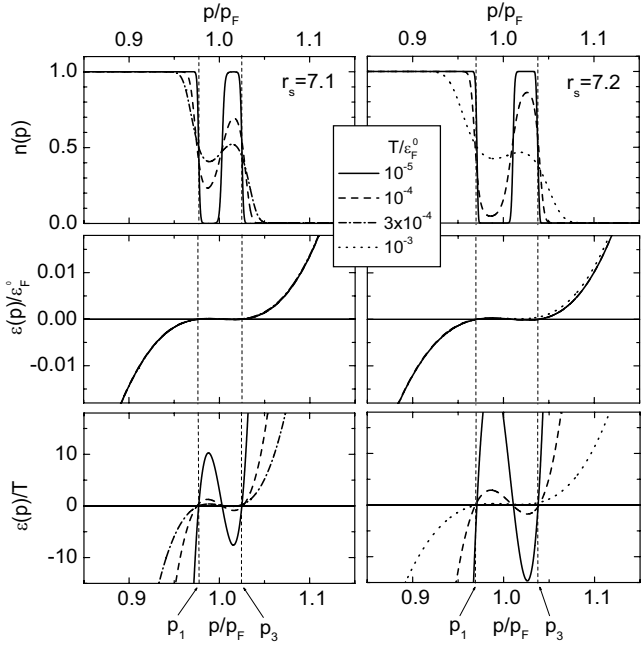}
\caption {Occupation number $n(p)$ (top panels),
single-particle spectrum $\epsilon(p)$ in units of $\epsilon_F^0$
(middle panels), and ratio $\epsilon(p)/T$ (bottom panels) for the
interaction model (\ref{model_2pf}) at $r_s=7.1$ (left column) and
$r_s=7.2$ (right column), both exceeding the QCP value $r_{\infty}=7.0$.
Interaction parameters are the same as for Fig.~\ref{fig:eg_fl}.
All three quantities are plotted as functions of $p/p_F$ at different
temperatures (measured in units $\varepsilon_F^0$) below the
crossover temperature $T_m=10^{-3}\varepsilon^0_F$.} \label{fig:eg_b}
\end{figure}

\begin{figure}[t]
\includegraphics[width=1.0\linewidth,height=1.\linewidth]{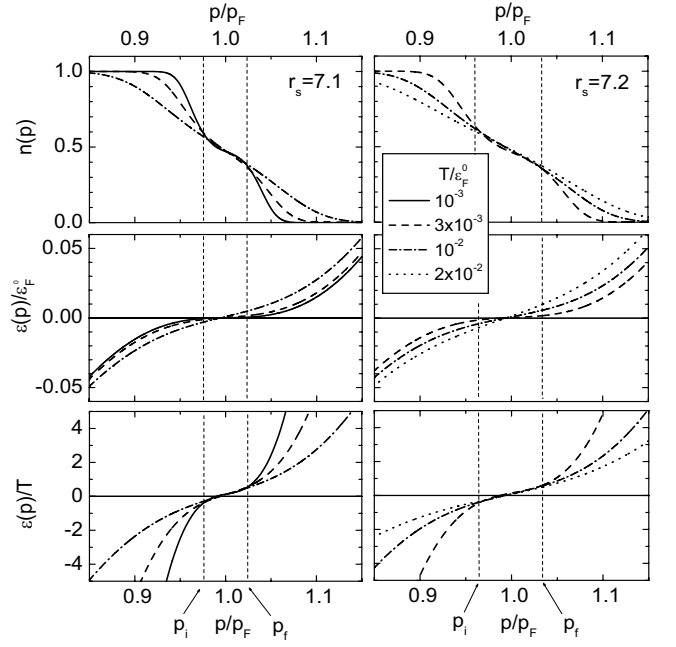}
\caption {Same as in Fig.~\ref{fig:eg_b} but at
temperatures {\it above} $T_m$.} \label{fig:eg_fc}
\end{figure}

\begin{figure}[t]
\includegraphics[width=0.7\linewidth,height=1.18\linewidth]{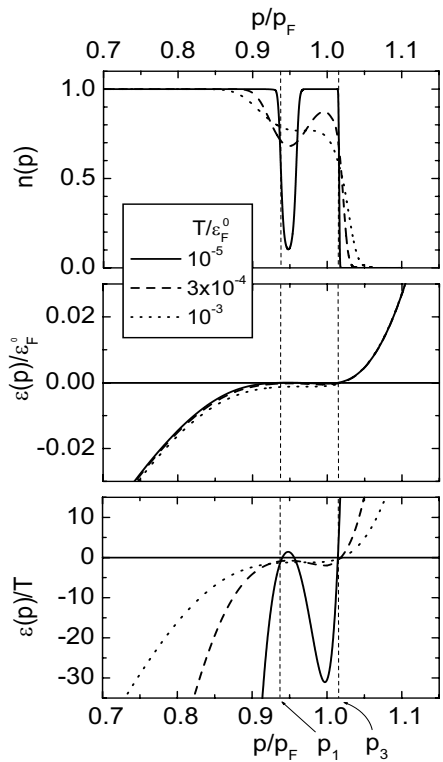}
\caption{Occupation numbers $n(p)$ (top panel),
single-particle spectrum $\epsilon(p)$ in units of
$10^{-3}\,\varepsilon_F^0$ (middle panel), and ratio
$\epsilon(p)/T$ (bottom panel), plotted versus $p/p_F$ at three
line-type-coded temperatures in units of $\varepsilon_F^0$, taken
below the transition temperature $T_m=3\times
10^{-3}\,\varepsilon^0_F$. Interaction model (\ref{model_0pf}) is assumed
with parameters $g_3=0.45$ and $\beta_3=0.07$.} \label{fig:b_0pf}
\end{figure}

\begin{figure}[t]
\includegraphics[width=0.7\linewidth,height=1.18\linewidth]{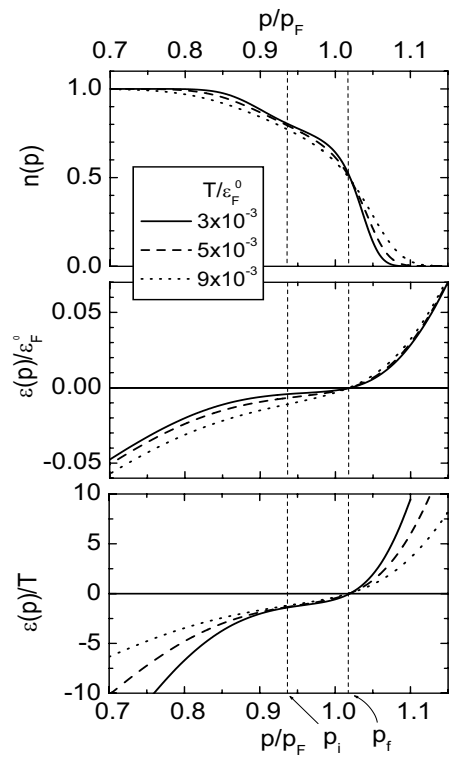}
 \caption {Same as in Fig.~\ref{fig:b_0pf} but at
temperatures {\it above} $T_m$.}
 \label{fig:fc_0pf}
\end{figure}

In Figs.~\ref{fig:eg_fl}--\ref{fig:fc_0pf}, we show results from
numerical calculations of the spectrum $\epsilon(p,T)$ based
on Eq.~(\ref{lansp}), for two different interaction functions $f$
that are regular in momentum space.  The choice of the first of these,
\beq
f(k)=-g_2{\pi\over M} {1\over (k^2/4p_F^2-1)^2+\beta^2_2} \ ,
\label{model_2pf}
\eeq
is motivated by the fact that for the 2D electron gas, it
can adequately describe the results of microscopic calculations\cite{bz}
of the zero-temperature single-particle spectra $\epsilon(p,T=0)$
on the FL side of the corresponding QCP, i.e.\ at $r_s<r_{\infty}=7.0$.
(Again, the dimensionless parameter $r_s $ is the radius of the
volume per particle measured in units of the atomic Bohr radius,
thus related to the Fermi momentum by $r_s = \sqrt{2}Me^2/p_F$.)
The second interaction,
\beq
f(k)= g_3{\pi^2p_F\over M}{1\over k^2+\beta^2_3p^2_F}\ ,
\label{model_0pf}
\eeq
which is relevant to a 3D system, is chosen because it was employed
in the first work\cite{zb} addressing the deep connections
between the two types of the topological phase transitions
considered in the present article.
The results of calculations with these two interactions are
compared with those obtained from analytic solution, at $T=0$,
of a model of fermion condensation in a 3D having the
singular interaction function \cite{physrep}
\beq
f(k)=g_s{\pi^2\over M}{e^{-\beta_s k/p_F}\over k}    \  .
\label{model_yuka}
\eeq

Fig.~\ref{fig:eg_fl} displays the single-particle spectrum
$\epsilon(p)$ and the group velocity $d\epsilon(p)/dp$, as
calculated on the FL side of the QCP for the model corresponding to
the interaction (\ref{model_2pf}).  The group velocity behaves
as a parabolic function of momentum $p$.
With increasing $r_s$, the bottom of the parabola gradually moves
downward, and, when $r_s$ reaches $r_{\infty}$, makes contact with
the horizontal axis exactly at the Fermi momentum $p_F=p_{\infty}$.
Consequently, the spectrum $\epsilon(p,T=0)\sim (p-p_{\infty})^3$
has an inflection point at the Fermi surface.\cite{ckz}
When evaluated with
$n(p)=n_F(p)$, the group velocity changes sign from positive to
negative as $r_s$ passes $r_\infty$, in agreement with the
topological scenario for the QCP.\cite{ckz}  In the vicinity of the
QCP, the temperature dependence of the group velocity obeys the
relation $v_F(T)\sim T^{2/3}$, again in agreement with the result
obtained in Ref.~\onlinecite{ckz}.

Figs.~\ref{fig:eg_b} and \ref{fig:eg_fc} show numerical results for
the quasiparticle momentum distribution $n(p)$ and the spectrum
$\epsilon(p)$, as calculated for the interaction function (\ref{model_2pf})
both below and above the temperature $T_m$.  Inevitably,
the value of $T_m$ is somewhat uncertain, because
the alteration of the FL behavior in this temperature region is
associated with {\it crossover}, rather than with some second-order
phase transition.  Nevertheless, comparison of Fig.~\ref{fig:eg_b}
($T<T_m$) and Fig.~\ref{fig:eg_fc} ($T>T_m$) reveals striking
changes in the structure of both of the functions
$\epsilon(p)$ and $n(p)$. Indeed, as seen in the top
panel of Fig.~\ref{fig:eg_b}, a well-defined multi-connected Fermi
surface, witnessed by a pronounced gap in filling, exists only at
extremely low $T< 10^{-4}\varepsilon^0_F$, while at $T\simeq
T_m\simeq 10^{-3}\varepsilon^0_F$, the gap in the occupation
numbers closes.

On the other hand, upon inspection of the top panel of
Fig.~\ref{fig:eg_fc}, we observe that within a region
region $p_i<p<p_f$ all of the curves for $n(p,T)$ for $T > T_m$
collapse into a single one---i.e., the momentum distribution
$n(p)$ becomes $T$-independent.  This behavior persists until
rather high $T$.  Fig.~\ref{fig:eg_fc} also indicates that
the range $\eta=p_f-p_i$ practically coincides with the
difference-of-roots $\Delta p = p_3-p_1$ (cf.\ Subsec.~III.A),
so that in accordance with Eq.~(\ref{relq}) we can write
\beq
\eta \sim \sqrt{\rho_{\infty}-\rho}\  .
\label{etfc}
\eeq
Moreover, comparison of the bottom panels of Figs.~\ref{fig:eg_b}
and \ref{fig:eg_fc} demonstrates that the huge variations of the ratio
$\epsilon(p)/T$ that are so prominent at $T<T_m$ completely
disappear in the FC domain at $T>T_m$.  Thus, again we discover
{\it scaling behavior}: within the range $p_i < p < p_f$ all
of the curves representing the ratio $\epsilon(p)/T$ collapse
into a single one.

The same conclusions follow from parallel calculations carried
out for the interaction (\ref{model_0pf}).  For this case,
Figs.~\ref{fig:b_0pf} and \ref{fig:fc_0pf}, dedicated
respectively to $T< T_m$ and $T > T_m$, trace the behavior
with temperature of the spectrum $\epsilon(p)$ and momentum
distribution $n(p)$.

\begin{figure}[t]
\includegraphics[width=0.7\linewidth,height=1.18\linewidth]
{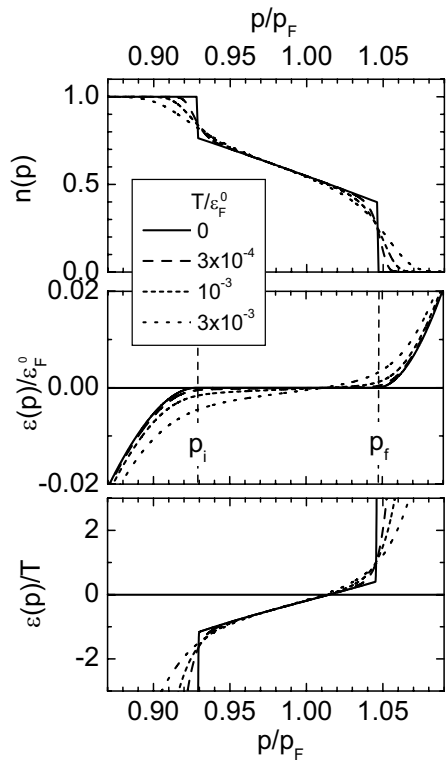}
 \caption {Occupation number $n(p)$ (top panels),
single-particle spectrum
$\epsilon(p)$ in units of $\varepsilon_F^0$ (middle panels),
and ratio $\epsilon(p)/T$,
(bottom panels) for the
model (\ref{model_yuka}) with the parameters $\beta=10$ and
$g_s=70.0$. All three quantities are shown as functions of $p/p_F$
at different line-type-coded temperatures measured in units of
$\varepsilon_F^0$.} \label{fig:fc_yuka}
\end{figure}

The scaling behavior observed in Figs.~\ref{fig:eg_b}--\ref{fig:fc_0pf}
is also evident in Fig.~\ref{fig:fc_yuka}, where corresponding results of
calculations based on Eq.~(\ref{lansp}) and the interaction
model (\ref{model_yuka}) are displayed.  The single noteworthy
difference is that the linear-in-$T$ dispersion of the single-particle
spectrum $\epsilon(p)$ characteristic of the FC domain is already
in effect at $T\to 0$, i.e., $T_m=0$ in this case.  The results of
similar numerical calculations for the singular interaction function
(\ref{cur}), to be published separately, support the same general
conclusions, in particular with respect to scaling behavior.
From the evidence gathered for different singular interaction
functions, we infer that scaling features that govern the
flattening of single-particle beyond the QCP and at $T > T_m$
are {\it universal}.  The specifics of the interaction function
affect only the value of the crossover temperature $T_m$.

\section{Relevance of the QCP topological scenario to reality}

Here we provide a brief assessment of the relevance of the theory
outlined in this article to the contemporary experimental situation.
At the outset we should emphasize that this theory is applicable
only to systems whose properties are aptly described by the
conventional FL approach on the weakly correlated side of the
QCP.  If this condition is not met, the theory is irrelevant.

At present, experimental information on the QCP is available for
three types of strongly correlated Fermi systems. The first type
is a silicon inversion layer in MOSFET, with the electrons
forming a homogeneous 2D liquid.  It is known from measurements
of the magnetic susceptibility and the Shubnikov-de Haas oscillations
\cite{sarachik,pudalov,shashrev,dolgrev} that the 2D electron
gas does obey standard Landau theory on the FL side of the QCP.
Furthermore, recent measurements \cite{shashprb} carried out
for two different versions of the crystal structure, namely $(100)$ and
$(111)$, and {\it for differing disorder}
have nevertheless demonstrated that the density at which the effective
mass $M^*$ diverges remains the same,
$\rho_{\infty}\simeq 0.8\times 10^{11}$\,cm$^{-2}$.  This finding
implies that {\it electron-electron interactions rather than disorder}
are responsible for the occurrence of the QCP in the silicon inversion
layer.  Moreover, the experimental value of the Stoner factor, which
specifies the proximity to the ferromagnetic phase transition,
is shown not to be enhanced relative to its value in the volume of
the silicon crystal.\cite{shashrev}  This finding implies that
ferromagnetism is irrelevant to the rearrangement of the ground
state that takes place at the QCP.  From the theoretical side,
we have already mentioned that results\cite{bz} from microscopic
calculations bearing on the divergence of effective mass of
the 2D electron gas are in agreement with the topological scenario for
the QCP.  All of the above factors, both empirical and computational,
speak for the relevance of the proposed topological scenario to
the reality of the 2D electron gas in the vicinity of the QCP.

Salient experimental information is also available for a second
class of Fermi systems having a QCP: films of $^3$He atoms lying
on different substrates.  With certain qualifications, these
systems may be modeled as 2D liquid $^3$He.  If the dimensionless
Landau parameters specifying the effective interaction between
$^3$He atoms are rather small, the laboratory systems may found to
obey standard FL theory.  There is experimental evidence that this
is indeed the case\cite{godfrin,saunders,sscience}---in spite of
the violation of the Galilean invariance and the presence of
disorder, effects introduced by the substrate.  As reflections of
a divergent density of states, the presence of the QCP in $^3$He
films is exhibited in measurements of the spin susceptibility
$\chi$,\cite{godfrin,saunders} and the specific heat
$C$,\cite{saunders,sscience} with the quantitative results showing
a substantial dependence on the nature of the substrate.  A
crucial feature of the measurements\cite{godfrin,sscience} is the
curious Curie-like behavior\cite{godfrin,sscience} found for the
spin susceptibility: $\chi(T,\rho)=C_{\mbox{\scriptsize
eff}}(\rho)/T$, with the effective Curie constant
$C_{\mbox{\scriptsize eff}}$ depending strongly on the density
$\rho$ in the QCP region.  The challenge of explaining this
experimentally documented {\it crossover} from Pauli-like to
Curie-like behavior is still open. A microscopic understanding is
necessarily complicated by interactions between substrate and
helium atoms and by layering effects at higher areal densities. In
the next section we will see that the topological scenario for the
QCP shows promise of capturing the essential physics of the
crossover.

The third and most extensive class of systems possessing a QCP
is that of the heavy-fermion metals, in which the crystal
lattice gives rise to anisotropy of the Fermi surface.  We
have argued in Subsec.~II.B.2 that short-range antiferromagnetic
fluctuations cannot be relevant to the QCP in these systems.  On the
other hand, the existence of topological phase transitions in this
class of systems is well known,\cite{lifshitz,abr} as pointed out
in the same subsection.
{\it A priori} there appear to be no
serious obstacles to implementation of the topological QCP
scenario for the heavy-fermion metals.

\section{Unconventional thermodynamics beyond the QCP}

Theoretical results based on the ideas and method we have
developed allow one to reproduce available experimental data
on properties of several concrete strongly-correlated Fermi systems
in the immediate vicinity of the QCP (for example, see
Refs.~\onlinecite{ckz,yak,kz2007}).  However, such
validation is to some extent inconclusive, since as a
rule there exist at least two alternative models that
are capable of explaining the data at the same level
of accuracy.  Accordingly, in this section we will focus
on certain unusual features of the thermodynamic properties
of strongly correlated systems beyond the QCP that emerge
in the proposed topological scenario.  Detailed comparisons
of the theory with experimental data is reserved for a
future article.

\subsection{Entropy excess}

In standard FL theory where the entropy $S$ is given by Eq.~(\ref{entr}),
the curve $S(T)$ starts at the origin and begins to rise linearly with
$T$.  The situation changes at the QCP.  In the conventional scenario,
which is predicated on a vanishing quasiparticle $z$-factor, one has
$S(T)\sim C(T)\sim T\ln (1/T)$. In the topological scenario for the QCP,
the divergence of the ratio $C(T)/T$ is even stronger:\cite{ckz}
$C(T)/T\sim T^{-2/3}$. The standard FL behavior of $C(T)/T$ does
re-emerge beyond the QCP, where the density of states is again finite.
However, recovery of this behavior occurs only at extremely low
temperatures, with the proviso that $T < T_m$.
At higher $T$, the FC forms in the domain $\Omega$, and its
characteristic NFL momentum distribution, given by $n_*(p)$, leads
to a drastic change in the behavior of the entropy
$S(T)$.\cite{ks,physrep,zk4,yak}
The basic entropy formula (\ref{entr}) of the original quasiparticle
formalism remains intact, but due to the NFL component in $n_*(p)$,
the system is seen to possess a {\it $T$-independent} entropy
excess $S_*(\rho)$.  Its value does not depend of the manner of
its evaluation, since Eq.~(\ref{kolm}) and (\ref{entr}) provide
{\it the same result}.

The situation we now face---with the strongly correlated
fermion system having a finite value $S_*$ of the entropy at
$T=0$---resembles that encountered in a system of localized spins.
In the spin system, the entropy referred to one spin is simply
$\ln2$, while in the system having a FC, we have
$S_*/N\simeq\eta \ln2$, where $\eta=\rho_f/\rho$ is the
dimensionless FC parameter.

Numerical calculations demonstrate that within the domain $\Omega$,
the momentum distribution $n_*(p)$ changes rapidly under variation of the
total density $\rho$.  The corresponding nonzero value of the derivative
$\partial S_*/\partial\rho$ produces a huge enhancement of the thermal
expansion coefficient $\beta\sim\partial S(T)/\partial\rho\simeq \eta$
with respect to its FL value, proportional to $T$.\cite{zksb}
Consonant with this result, experimental data\cite{oeschler} show
that at low $T$ in many heavy-fermion metals, $\beta$ is indeed almost
temperature-independent and exceeds typical values for ordinary
metals by a factor $10^3$--$10^4$.  To our knowledge,
no theory has previously been advanced to explain
this enhancement.

\subsection{ Curie-Weiss behavior of the spin susceptibility}

Another peculiar feature of  strongly correlated Fermi systems beyond the
QCP involves the temperature dependence of the spin susceptibility
$\chi(T)=\chi_0(T)/(1+g_0\Pi_0(T))$,
where
\beq
\chi_0=\mu^2_e\Pi_0(T)=-2\mu^2_e\int {dn(p,T)\over d\epsilon(p)}d\upsilon
\,,
\label{chio}
\eeq
and $g_0$ is the spin-spin component of the interaction function
(which remains unchanged through the critical density region).
As mentioned before, beyond the QCP the standard Pauli behavior
of $\chi(T)$ prevails only at $T<T_m$, and its value, proportional
to the zero-temperature density of states
$N(0)\sim |\rho-\rho_{\infty}|^{-1}$, turns out to be
greatly enhanced.

At $T>T_m$, insertion of $n_*(p)$ into Eq.~(\ref{chio}) yields
the Curie-like term
\beq
\chi_*(T)=\mu^2_e{ C_{\mbox{\scriptsize eff}}(\rho)\over T} \,,
\eeq
with an effective Curie constant
\beq
C_{\mbox{\scriptsize eff}}(\rho)=2\int n_*(p)(1-n_*(p))d\upsilon
\label{ceff}
\eeq
that depends dramatically on the density.\cite{zk4,yak}
Since  $C_{\mbox{\scriptsize eff}}$ is proportional to
the FC parameter $\eta$, we infer that
\beq
C_{\mbox{\scriptsize eff}}\simeq S_*  \ .
\eeq
Thus, all compounds in which the spin susceptibility exhibits the
Curie-like behavior possess a large entropy. Furthermore, in the
whole temperature interval from $T=0$ to $T>T_m$, the spin susceptibility
of a Fermi system beyond the QCP possesses Curie-Weiss-like behavior
$\chi(T)\sim 1/(T-T_W)$ with a {\it negative} Weiss temperature $T_W$.
Measurements in $^3$He films on various substrates\cite{godfrin,sscience}
and in numerous heavy-fermion compounds
(cf.\ Refs.~\onlinecite{steglich3,tayama}) provide examples of
this NFL behavior. We emphasize that in our scenario, the negative
sign of $T_W$ holds {\it independently} of the character
of the spin-spin interaction, repulsive or attractive.
(In the latter case, the magnitude of this interaction must
not exceed certain limits, as indicated below in
Subsec.~VIII.B.)  At the same time, in a
system of localized spins the Weiss temperature has a negative
sign only if the spin-spin interaction is {\it repulsive}.
In this case, however, the Stoner factor must be suppressed.
The necessity of reconciling the negative sign of $T_W$
with the enhanced Stoner factor observed experimentally in
the vicinity of the QCP\cite{steglich,gegenwart} creates
insurmountable difficulties for the standard collective
QCP scenario.

Another conspicuous feature of the physics beyond the
QCP is associated with the Sommerfeld-Wilson ratio
$R_{SW}=T\chi(T)/\mu^2_eC(T)\sim S_*/C(T)$. Since
the excess entropy $S_*$ does not depend on $T$, it
makes no contribution to the specific heat $C(T)$;
consequently one should see a great enhancement of
$R_{SW}$.

\section{Release of entropy stored in the fermion condensate}
The diversity of phase transitions occurring at low
temperatures is one of the most spectacular features of
the physics of many heavy-fermion compounds.  Within the
standard collective scenario,\cite{hertz,millis} it is hard
to understand why these transitions are so different from
one another and their critical temperatures are so
extremely small.  However, such diversity is endemic to systems
with a FC.  Its source may be traced to
an obvious fact: The existence of the excess entropy $S_*$ at
$T=0$ would contradict the third law of thermodynamics (the
Nernst theorem).  We may recall that in order to relieve themselves
of excess of the entropy, systems of localized spins order magnetically due to
spin-spin interactions.  The situation in systems with a FC is
similar, but there are many ways to release the excess entropy
$S_*$ as $T$ goes down to zero.  One possible route for eliminating
$S_*$, already considered, is the crossover between the state with a
FC and a state with a multi-connected Fermi surface as the
temperature drops below $T_m$.  However, this is by no means a
unique prescription for removing the entropy excess.  Other
options are associated with second-order phase transitions,
involving violation of a symmetry of the ground state.

\subsection{Superconducting phase transitions.}
It is instructive to begin with superconducting phase
transitions, considered already in the first article\cite{ks}
devoted to fermion condensation.  A necessary condition for
a superconducting transition to come into play is that
its critical temperature $T_c$ exceeds $T_m$.  To determine
$T_c$, one sets $\Delta_L\to 0$ in the well-known BCS
gap equation, yielding
\beq
\Delta_L(p,T\to 0)=\int {\cal V}_L(p,p_1){\tanh[ \epsilon(p_1)/2T]
\over 2\epsilon(p_1)}
\Delta_L(p_1,T) d\upsilon_1  \,,
\label{bcs}
\eeq
where ${\cal V}_L$ is the effective interaction between
quasiparticles in the $L$-wave pairing channel.  It can be
demonstrated that the FC contribution to the integrand on
the right-hand side of this equation is dominant.  After
some algebra employing the identity
\beq
{{\tanh \left( \epsilon(p)/2T \right) }\over {\epsilon(p)}}
={1-2n_*(p)\over T\ln \left[(1-n_*(p))/n_*(p)\right]}\ ,\quad {\bf
p} \in \Omega \ ,
\eeq
one obtains
\beq
T_c={\cal V}_L(p_F,p_F)\int_{\Omega} {1-2n_*(p)\over 2\ln
\left[(1-n_*(p))/n_*(p)\right]} d\upsilon \ .
\label{bcs1}
\eeq
Confining the integration to the small FC domain, one arrive
at\cite{ks}
\beq
T_c\simeq  \lambda\eta\varepsilon^0_F \,,
\label{tcfc}
\eeq
where $\lambda$ is a dimensionless pairing constant and
$\eta\sim |\rho-\rho_{\infty}|^{1/2}$ is the FC parameter.
We see now that a remarkable situation arises if the pairing
constant is large enough to ensure satisfaction of $T_c>T_m$.
In contrast to the exponentially small BCS critical temperature
$T_c\sim e^{-2/\lambda}$, the critical temperature in a
system having a FC turns out to be a {\it linear} function
of $\lambda$.

\subsection{Phase transitions in the particle-hole channel}
Along the same lines, we can consider the possibility of a
collapse of particle-hole collective modes in systems with
a FC.
Here we restrict ourselves to long-wave transitions that
 give rise to a deformation of the Fermi surface.
Manipulations similar those previously applied lead to
the relation
\beq
 O_L( p)=-\int {\cal F}_L( p, p_1) {dn(p_1)\over d\epsilon(p_1)}
O_L( p_1) d\upsilon_1  \ ,
\label{shape}
\eeq
where $O_L(p)$ is shape function characterizing the deformation.
In the FC region we have $dn(p)/d\epsilon(p)=n_*( p)(1-n_*( p))/T$,
so that upon retaining only the FC contribution Eq.~(\ref{shape})
is recast as an equation for the transition temperature,
\beq
T_N=-{\cal F}_L( p_F, p_F)\int_{\Omega}  n_*( p) (1-n_*( p))
d\upsilon\simeq -f_L\eta\varepsilon^0_F \ .
\eeq
On observing that the dimensionless Landau parameters
$F_L=f_L/N(0)$ keep their values through the critical region, we
may express the transition temperature in the form
$T_N=F_L\eta/N(0)$. Remembering that the FC density $\eta$ is
proportional to $|\rho-\rho_{\infty}|^{1/2}$, while the density of
states behaves as $N(0)\sim |\rho-\rho_{\infty}|^{-1}$, the
estimated value of $T_N$ appears to be proportional to
$|\rho-\rho_{\infty}|^{3/2}$, being comparable with the crossover
temperature $T_m$. In the case of ferromagnetism, the condition
$T_N>T_m$ is met only if the spin-spin component of the
interaction function is negative and sufficiently large.
Otherwise, the system avoids the ferromagnetic phase transition
and the spin susceptibility obeys the Curie-Weiss law $\chi(T)\sim
1/(T-T_W)$ with a negative Weiss temperature $T_W$.

Remarkably, in either of the above candidates posed as a
mechanism for release of the excess entropy $S_*$
through a symmetry-breaking second-order phase transition,
the unconventional state with a FC always corresponds to the
{\it high-temperature phase}, while the low-temperature phase
possesses more familiar properties.  Phase transitions occur
in any channel where the sign of the effective interaction
is suitable for the transition, {\it provided} the value of
the effective coupling constant is sufficient to produce
the inequality $T_c>T_m$.  Otherwise, the entropy excess
is released through the crossover leading to formation of a
multi-connected Fermi surface.  We see, then, that the
diversity of phase transitions in the topological scenario
for the QCP is due to the accumulation by the FC of a big
entropy at extremely low temperatures.  Because of the smallness
of the parameter $|\rho-\rho_{\infty}|$, the transition temperatures
appear to be very low and the phase transitions themselves
inevitably of quantum origin.  Several phases specified by
different order parameters can in principle coexist with each
other, giving rise to states of an intricate nature.  As the
temperature decreases to zero, these phases can replace one another.

\section{Conclusion}

A basic postulate of standard Fermi liquid theory reads: ``At the
transition from a Fermi gas to a Fermi liquid, the classification
of energy levels remains unchanged'' (Ref.~\onlinecite{lanl}, p.\ 11).
Some 20 years ago, the first paper on the theory of fermion
condensation\cite{ks} demonstrated that this postulate is
incorrect beyond a quantum critical point, while maintaining
the essence of the quasiparticle picture.
The paper triggered a wave of criticism and disbelief;
the judgment, ``This theory is an artifact of the Hartree-Fock
method,'' was typical.  By now, debates on the subject have
become pointless: numerical calculations based on Eq.~(\ref{lansp}),
carried out during the last decade and discussed in the
present work, provide the best way to answer critics.

It has been a principal goal of this article to investigate the
structure of the Fermi surface beyond the quantum critical
point within the original Landau quasiparticle pattern.\cite{lan}
We have shown that at $T=0$ there are two different realizations
of this pattern, both of which involve a topological phase
transition.  The first is associated with the emergence of a
multi-connected Fermi surface. The second entails the formation
of a fermion condensate, implicated by the emergence of completely
flat portion of the single-particle spectrum $\epsilon(p)$ that
may be envisioned as a virtual swelling of the Fermi surface.
Such an inflation of the Fermi surface can occur if the
Landau interaction function $f$ contains components of
long range in coordinate space.

We have performed a series of numerical calculations that
serve to demonstrate the existence of a crossover between
the two types of topological structure at extremely low
temperature, an effect that introduces a new energy scale
$T_m$ into the problem.   At $T > T_m$, both the momentum
distribution $n(p)$ and the single-particle spectrum
$\epsilon(p)$ exhibit universal features inherent in strongly
correlated Fermi systems beyond the QCP.  To be specific,
scaling features pertaining to the behavior of the momentum
distribution $n(p)$ and the (scaled) single-particle energy
$\epsilon(p)/T$ in the vicinity of the Fermi surface are
independent of the assumed form of the interaction function
$f$ (and notably whether on not it contains long-range
components).  The choice of $f$ does, however, affect
the value of $T_m$.

We have explored intriguing features of Poincar\'e mapping
as a technique for iterative solution of the nonlinear integral
equation (\ref{lansp}) that connects the group velocity and the
quasiparticle momentum distribution at zero temperature.  For
regular interaction functions, the sequence of iterative maps
converges to the true solution beyond the quantum critical point
and describe a new ground state with a multi-connected Fermi
surface.  However, if the Landau interaction function $f$
has a component of long range in coordinate space, 2-cycles
are generated in the standard iteration procedure, a behavior
symptomatic of the inadequacy of this solution algorithm
as well as the failure of standard Fermi liquid theory
beyond the quantum critical point.
Adopting a refined iteration
procedure (in which the input for the next step is a mixture
of the outputs of the two preceding iterations), the spurious
2-cycles no longer arise.  However, the new sequence of
iterative maps acquires chaotic features in a region
of momentum space adjacent to the Fermi surface that
{\it coincides} with the domain in which the 2-cycles
were previously found.

We have elaborated a special procedure for averaging the successive
outputs of the iteration process and demonstrated that the averaged
single-particle energies and occupation numbers so obtained coincide
with those inherent in states with a fermion condensate.
The exceptional states involving a fermion condensate are shown
to possess a nonzero entropy at zero temperature.  This result
does not contradict basic laws of statistical physics
{\it if and only if} the ground state is degenerate.  In effect,
the comprehensive analysis of Eq.~(\ref{lansp}) presented here
affirms the consistency of the properties of states possessing
a fermion condensate.  The new insight into the role of chaos
theory in the quantum many-body problem that has been gained
through our analysis lays a basis for future studies with
the potential of wide implications.

Finally, we have investigated pathways for releasing the entropy
excess stored in the fermion condensate that are associated
with different {\it quantum} phase transitions---necessarily
of quantum origin since the transitions involved occur at
extremely low temperatures.

\section*{Acknowledgments}

We thank G.~Baym, V.~Dolgopolov, M.~Norman, Z.~Nussinov, J.~Nyeki,
S.~Pankratov, J.~Saunders, V.~Shaginyan, A.~Shashkin, F.~Steglich,
J.~Thompson, G.~Volovik, V.~Yakovenko, and N.~Zein
for valuable discussions.  This research was
supported by the McDonnell Center for the Space Sciences, by
Grant No.~NS-3004.2008.2 from the Russian Ministry of Education
and Science, and by Grants Nos.~06-02-17171 and 07-02-00553 from
the Russian Foundation for Basic Research.  JWC is grateful
to Complexo Interdisciplinar of the University of Lisbon and
to the Department of Physics of the Technical University of
Lisbon for gracious hospitality during a sabbatical leave; and
to Funda\c{c}\~{a}o para a Ci\^{e}ncia e a Tecnologia of the
Portuguese Minist\'erio da Ci\^{e}ncia, Tecnologia e Ensino
Superior as well as Funda\c{c}\~{a}o Luso-Americana for
research support during the same period.

\end{document}